\documentclass[prl,showpacs,notitlepage,twocolumn,superscriptaddress]{revtex4-1}

\pdfoutput=1
\usepackage{tabularx}
\usepackage{blindtext}
\usepackage{svg}
\usepackage{float}
\usepackage{mwe}
\usepackage{multirow}
\usepackage[colorlinks,linkcolor={blue},citecolor={blue},urlcolor={blue}]{hyperref}

\usepackage{graphicx}

\usepackage{bm,amsmath}

\usepackage{latexsym}

\usepackage{verbatim}

\usepackage{xcolor}
\usepackage{subfigure}

\usepackage{gensymb}

\usepackage{soul}

\usepackage{tikz}

\renewcommand*\vec[1]{\mathbf{\bm{#1}}}
\begin{document}

 \title[]{Effects of inertia on conformation and dynamics of tangentially-driven active filaments}
\author{Mohammad Fazelzadeh}
\affiliation{Institute of Physics,  University of Amsterdam, Amsterdam, The Netherlands
}%

\author{Ehsan Irani}
\affiliation{Institute for Theoretical Physics, Georg-August University of Göttingen, Friedrich-Hund Platz 1, 37077 Göttingen, Germany
}%

\author{Zahra Mokhtari}
\affiliation{Department of Mathematics and Computer Science
Institute of Mathematics, Free University of Berlin
}%
\author{Sara Jabbari-Farouji $^{*}$}
\affiliation{Institute of Physics,  University of Amsterdam, Amsterdam, The Netherlands
}%
 \email{Correspondence to: s.jabbarifarouji@uva.nl}


 \date{\today}

 \date{\today}

  \begin{abstract}
  
Active filament-like systems propelling along their backbone exist across the scales ranging from  motor-driven bio-filaments to worms and robotic chains. In macroscopic active filaments such as chain of robots,  in contrast to their microscopic counterparts, inertial effects on their motion cannot be ignored. Nonetheless, consequences of interplay between inertia and flexibility on shape and dynamics of active filaments remain unexplored. Here, we examine inertial effects on  flexible tangentially-driven active polymer model pertinent to above examples and we determine the conditions under which inertia becomes important. Performing Langevin dynamics simulations of  active polymers with underdamped and overdamped dynamics for a wide range of contour lengths and activities, we uncover striking inertial effects on conformation and dynamics at high activities. Inertial collisions increase the persistence length of active polymers and remarkably alter their scaling behavior. In stark contrast to passive polymers,  inertia leaves its fingerprint at  long times  by an enhanced  diffusion of the center of mass.  We rationalize inertia-induced enhanced dynamics by analytical calculations of center of mass velocity correlations, \emph{applicable to any active polymer model}, which reveal significant contributions from  active force fluctuations convoluted by inertial relaxation.
  
  \end{abstract}


\maketitle
\section{ Introduction} 
Active matter systems, consisting of self-driven units,  exhibit emergent properties which defy the laws of equilibrium statistical mechanics~\cite{Ramaswamy2010,Marchetti1}.  The majority of recent studies  have focused on   active particles moving in the realm of low Reynolds numbers, {\it e.g.} bacteria and active colloids whose motion is overdamped~\cite{Marchetti1,ABP-review}.
However, a wide range of macroscopic organisms including birds, fish, snakes, as well as synthesized agents such as microflyers~\cite{Scholz_18,3Dmicroflyer,lowen2020} and shaken granulate chains~\cite{granular_chain} often have elongated flexible shapes with  non-negligible masses which need to be accounted for description of their dynamics.
  Indeed, recent studies have revealed significant inertial effects on  dynamics of individual and collectives of isotropic and rigid active particles \cite{lowen2020, Scholz_18, Malakar, jahanshahi,inertial_flock_21,Inertial_AOUP,ABP_diffusion,trappped_ABP_inertia,Dai_inertia,Ineria_time_dependent,lowen-AOUP,caprini2022role}. Notably, inertia enhances diffusive dynamics of active Brownian particles~\cite{Scholz_18,lowen2020,lowen-AOUP}, unlike their passive counterparts which do not hold any memory of inertia at long times.
  These findings raise the interesting question whether inertia induces memory effects and enhanced dynamics for active macroscopic systems with flexible bodies like fishes and snakes.  \\ 
  %
   \begin{figure}[t]
    \centering
    \begin{tikzpicture}
            \draw (0,0) node[inner sep=0]{\includegraphics[width=\linewidth ]{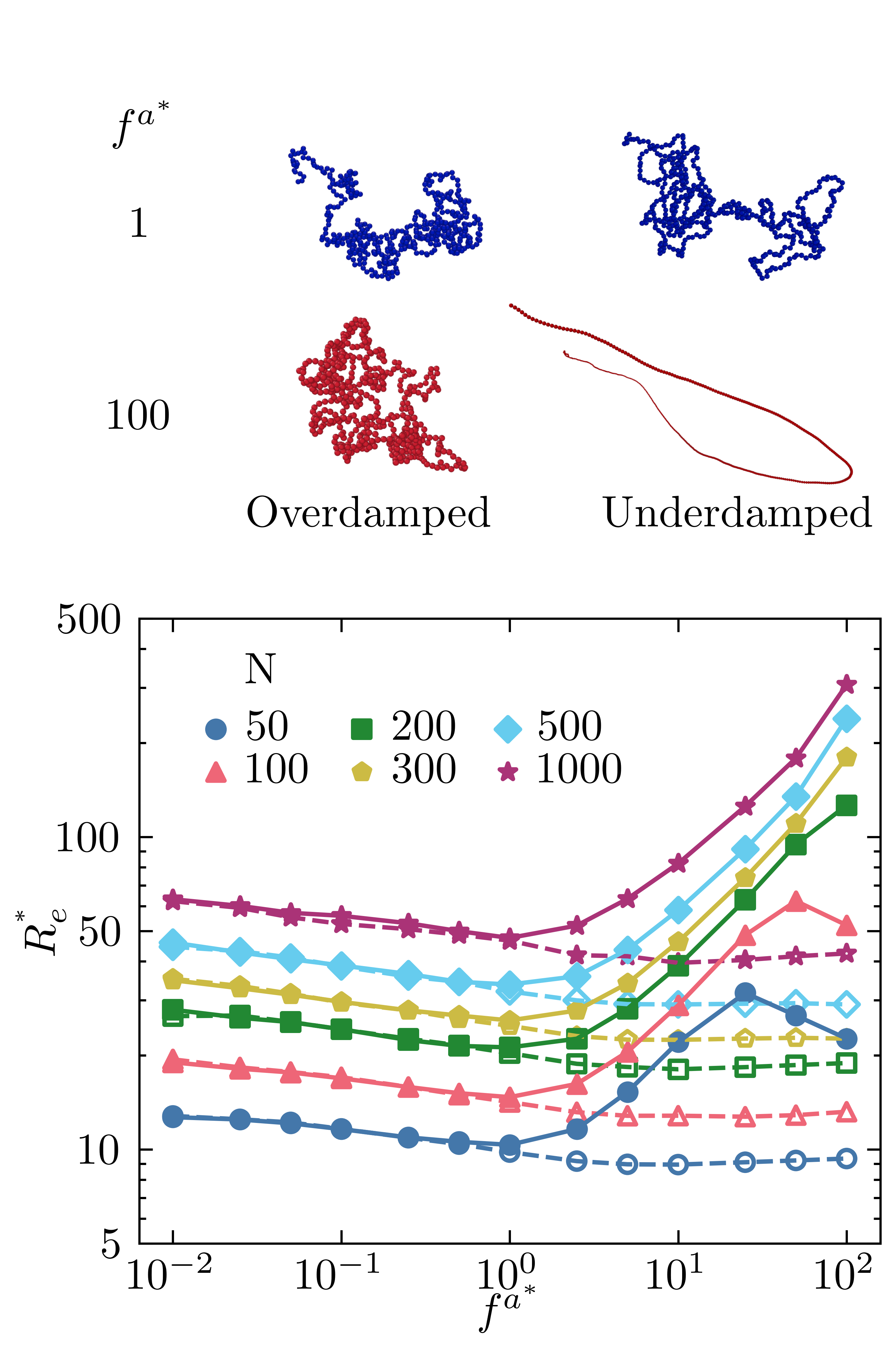}};
            \draw (3.3,5.95) node {\textbf{(a)}};
            \draw (3.3,-0.25) node {\textbf{(b)}};
            
    \end{tikzpicture}
    \vspace{-7mm}
    \caption{(a) Snapshots of active polymers of size $N=500$ with active forces $f^{a^{*}}=1$ (blue) and  100 (red)  in the steady state obtained in the overdamped (left) and underdamped (right) regimes. (b) The end-to-end distance $R_e^{ ^{*}}$ versus active force $f^{a^{*}}$ for different chain lengths  as given in the legend. The bold and empty symbols in panel (b) correspond to underdamped and overdamped chains, respectively. The lines are guides to the eyes.}  
    \label{fig:conformation}
    \vspace{-5mm}
\end{figure}

Presently, consequences of inertia on shape and dynamics of macroscopic active systems remain an open question. To explore the  role of inertia on flexible active particles, we consider the  widely studied system of linear active filaments~\cite{active_poly_review,active_pol_20} spanning a diverse range of biological systems such as biopolymers driven by molecular motors~\cite{biopolymers1,biopolymers2}, worms~\cite{aquarium_worms,worms_rheology,California_worms} and snakes~\cite{snakes}. Recently even filamentous robots~\cite{California_worms,snakes,Corentin} have been achieved, belonging to this class of active systems. 
Investigations of single overdamped active filaments~\cite{active_poly_review,active_pol_20,Isele,Active_flexible,active_priodic_force,Mokhtari,Ghosh_22} have revealed that interplay between activity and flexibility  profoundly alters the chain conformation and dynamics and it leads to an activity-dependent relaxation time~\cite{active_poly_review,ABPO_ANALYTICS-eisenstecken_gompper_winkler_2016,Active_flexible,Phillipps_22}.  
It also appears that details of the coupling between the local active force and the conformation of the polymer backbone are crucial for determining the overall dynamics of the polymer. However, it remains unknown under what conditions inertial effects become dominant as we change the scale and mass even if self-propulsion of filaments are governed by the same rules. 

Here, we examine the inertial effect on the  structural and dynamical features of \emph{tangentially-driven} active polymers~\cite{Isele}, in which the orientation of active force on each segment is parallel to the local tangent of the backbone. In this model, the total active force on the polymer is proportional to end-to-end vector,  thus the center of mass dynamics is directly coupled to the polymer conformation. We focus on this model as recent studies have shown that it provides a good description of several active filaments across the scales including biofilaments driven by molecular motors~\cite{Isele}, worms~\cite{nguyen2021emergent} and filamentous robots~\cite{Corentin}. Additionally, robotic snakes~\cite{snakerobot2,snakerobot1} with tangential activity have been realized. Keeping all active polymer model parameters identical, but varying inertia, we study the inertial effects as a function of  contour length and activity strength using Langevin dynamics simulations.

 Comparing the conformation and dynamics of active polymers in the overdamped and underdamped limits, we find remarkable inertial effects at sufficiently large activities. Inertial effects become non-negligible when  the timescale of advection by active force per unit of length becomes smaller than inertial timescale. At large activities, inertial collisions of active monomers with relative high velocities result in extended chain conformations, see Fig.~\ref{fig:conformation} (a), inducing an activity-dependent persistence length.  
  We find that  inertia enhances diffusive dynamics of the center of mass of  inertial  active polymers remarkably, in contrast to passive polymers  for which inertial effects vanish at long timescales ~\cite{inertia_Rouse,Deutsch-review,Deutsch}.
  To elucidate the origin of enhanced dynamics,  we put forward analytical calculations which derive  the center of mass velocity time autocorrelation function in the steady-state limit for an arbitrary active force distribution along a polymer backbone. Our calculations reveal that the enhanced long-time diffusion in any active polymer model stems from  fluctuations  of  total active force on a filament  convoluted by an exponential relaxation with inertial timescale. For tangentially-driven polymers which the active force is coupled to the polymer conformation, we show that the long-time diffusion coefficient is proportional to the  mean-squared end-to-end distance. Hence, our theory illuminates  the link between enhanced dynamics and extended polymer conformations.   
  
\section{Numerical simulations} To further elaborate our new insights, we start by outlining  our Langevin dynamics simulations of tangentially-driven active polymers.
The equation of motion for each monomer of an active chain of $N$ beads of  mass $m$  in three dimensions is described by  
\begin{equation}
\label{eq:Langevin}
m \ddot{\vec{r}}_i=-\gamma \dot{\vec{r}}_i- \sum_j \nabla_{\vec{r}_i} U(r_{ij})+\vec{f}^a_{i}+\vec{f}^{r}_{i}, 
 \end{equation}
 in which $\vec{r}_i$ is the coordinate  of bead $i$ with the dots denoting derivatives with respect to time, and $\gamma$ is the friction coefficient of the bead with the surrounding medium. $r_{ij}$ denotes the distance between beads $i$ and $j$.  The potential energy $U(r_{ij})$ includes contributions from harmonic springs of equilibrium length $\ell$ and stiffness $k_s$ between adjacent monomers  and interbead excluded volume interactions modelled by the WCA potential~\cite{WCA},
$ U_{\text{excl}}(r)=4  \epsilon \left[  (\frac{\sigma}{r})^{12} -(\frac{\sigma}{r})^6+\frac{1}{4}\right]$
for $ r< r_c=2^{1/6} \sigma$.
  $\vec{f}^a_{i}$  and $\vec{f}^{r}_{i}$ are the active and random forces acting on the bead $i$, respectively.  
 The  active force on each bead, except for end monomers, is given by:
 $\vec{f}^a_i=\frac{f^a}{2 \ell } (\vec{r}_{i-1,i}+\vec{r}_{i,i+1})$,  where $\vec{r}_{i,i+1}=\vec{r}_{i+1}-\vec{r}_{i}$ defines the bond vector connecting   $(i+1)$-th and $i$-th monomers. The active forces on the end monomers are given by $\vec{f}^a_1=\frac{f^a}{2\ell } \vec{r}_{1,2}$ and $\vec{f}^a_N=\frac{f^a}{2\ell } \vec{r}_{N-1,N}$. The spring constants are chosen very stiff   $k_s  \gg f^a/\ell$,  to ensure that the mean bond-length and polymer contour length remain almost constant,   see Supplemental Material (SM)~\cite{SI} for details. The random force is chosen as a white noise of zero mean and correlation $\langle \vec{f}^r_i(t) \cdot \vec{f}^r_j(t') \rangle=6 D_0 \gamma^2 \delta_{ij} \delta(t-t')$.

 We choose $\ell_u=\sigma$, $E_u=\epsilon$ and $\tau_u=\gamma \sigma^2/\epsilon$, with $\gamma=1$, as the units of  length, energy and time, respectively, denoting reduced quantities  with '$*$' superscripts.   We fix  $\ell/\sigma=1$ and dimensionless diffusion coefficient $D_0^*=0.1$. To elucidate the role of inertia, we compare the scaling behavior and overall dynamics of underdamped active polymers with $m^*=m \epsilon/(\gamma \sigma)^2=1$ and   overdamped chains $m^*=0$, varying chain length  $50\le N \le 1000$ and active force strength $f^{a^{*}}=\frac{f^a \sigma}{\epsilon}$ in the range $0.01 \le f^{a^{*}} \le 100$. Additionally, for chain length $N=500$ effects of varying mass  $0 \le m^* \le 5$ on polymer mean conformation and persistence length are also investigated  and included in Fig.~\ref{fig:lpMreM}.\\\\
 
 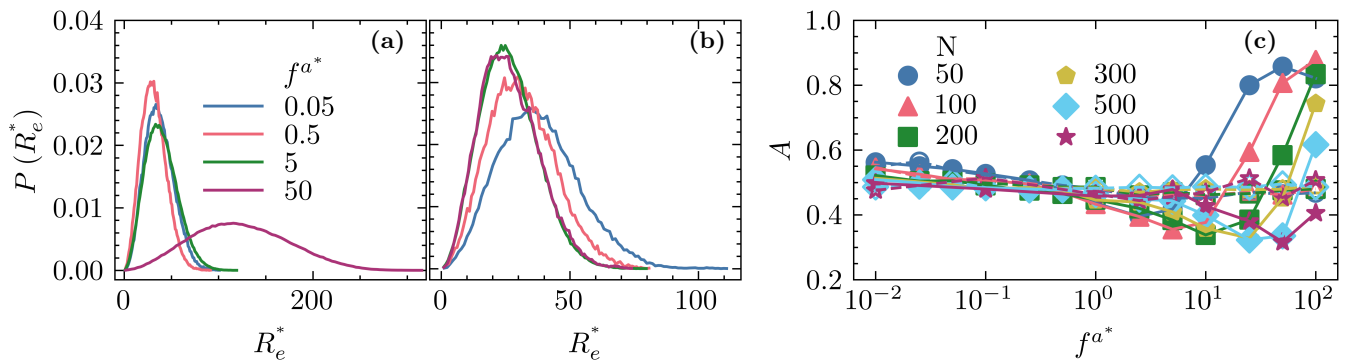
\begin{figure*}[ht]
    \centering
    \begin{tikzpicture}
            \draw (0,0) node[inner sep=0]{\includegraphics[width=\linewidth ]{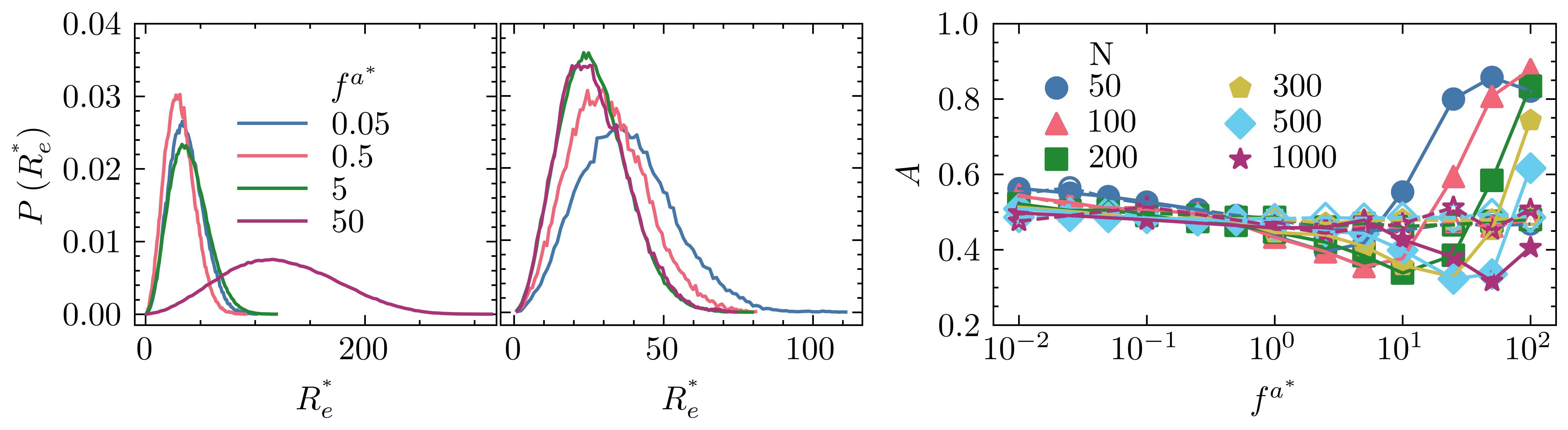}};
            \draw (-3.8,1.9) node {\textbf{(a)}};
            \draw (0.45,1.9) node {\textbf{(b)}};
            \draw (7.8,1.9) node {\textbf{(c)}};
    \end{tikzpicture}
    \vspace{-7mm}
    \caption{ PDF of end-to-end distance for active chains of $N=500$ at different activities  with (a) underdamped and  (b) overdamped dynamics. (c) Asphericity $A$ of active polymers of different chain lengths $50 \le N \le 1000$ as a function of $f^{a^{*}}$ in the underdamped (bold symbols) and overdamped (empty symbols) regimes. }
    \label{fig:PDF-ASPH}
    \vspace{-5mm} 
\end{figure*}

\section{Results} 
 \subsection{ Inertial effects on conformation}
We first investigate inertial effects on the global conformation of polymers. Fig.~\ref{fig:conformation}(a) shows that chain conformation is significantly more extended for high active forces in the underdamped scenario. To compare the mean conformation of polymers in the overdamped and underdamped regimes, we plot the
end-to-end distance  of active polymers,  obtained as $R_e^{ ^{*}}=  \sqrt{\langle \vec{R}^{2}_e \rangle}/\sigma$, in their steady-state as a function of activity $f^{a^{*}}$ for different chain lengths $N$ as presented in Fig.~\ref{fig:conformation}(b).
For $f^{a^{*}} \gg 1$, when  active force per monomer exceeds damping force, we observe a striking contrast between conformations of inertial and overdamped  active polymers.  Chains with underdamped dynamics swell, whereas overdamped chains slightly shrink upon increase of activity.
The onset of departure from the overdamped limit is set by the ratio of the inertial timescale $\tau_{\text{m}}= m/\gamma$  to  the time of advection by the active force per monomer  $\tau_{\text{adv}}= \sigma \gamma /f^a $. At large active forces when  $\tau_{\text{m}} / \tau_{\text{adv}}> 1$, large velocities gained by inertial  monomers promote collisions, resulting in chain unwinding.  Conformations of inertial chains at higher activities are more extended (see videos in SM~\cite{SI}),  resulting in broader probability distribution function (PDF) of end-to-end distance $P(R_{e}^ {^{*}})$, as can be inferred from Fig.~\ref{fig:PDF-ASPH} (a) and (b). For overdamped active chains, we observed a slightly more peaked $P(R_{e}^ {^{*}})$ at high activities,  reflecting a weak conformational shrinkage. This trend is similar to observation of reference~\cite{Active_flexible}, although the degree of shrinkage is much stronger in that work. This is because the model used in our simulations~\cite{Isele} is different from reference~\cite{Active_flexible} where a different tangential force rule is used~\cite{active_pol_20} and active forces on end beads are switched off. In contrast to the results of reference~\cite{Active_flexible}, we do not observe for polymers with overdamped dynamics a remarkable coil-globule-like conformational transition  upon increase of activity.

In addition, high-activity inertial chains have more anisotropy in their conformation as evidenced by their larger asphericity, which is defined as~\cite{asphericity}:
\begin{equation}
    A=\dfrac{\langle Tr^2 - 3M \rangle}{\langle Tr^2 \rangle}\;,
    \label{eq:asphericity}
\end{equation}
where $Tr=\lambda_1+\lambda_2+\lambda_3$, $M=\lambda_1\lambda_2+\lambda_2\lambda_3+\lambda_3\lambda_1$ and each $\lambda$ is an eigenvalue of the $3\times 3$ gyration tensor. Asphericity ranges from 0 for perfect spherical conformation to 1 for rod-like one. Fig.~\ref{fig:PDF-ASPH} (c) shows the asphericity $A$ of active polymers against $f^{a^{*}}$ for chains in the underdamped and overdamped regimes. While $A$  of inertial chains exhibits a striking increase for higher activities indicating an elongated shape, the asphericity of overdamped chains decreases slightly with $f^{a^{*}}$, akin to the trends observed in reference~\cite{Active_flexible}, albeit much weaker. 

  \begin{figure}[h]
    \centering
     \includegraphics[width=1\linewidth]{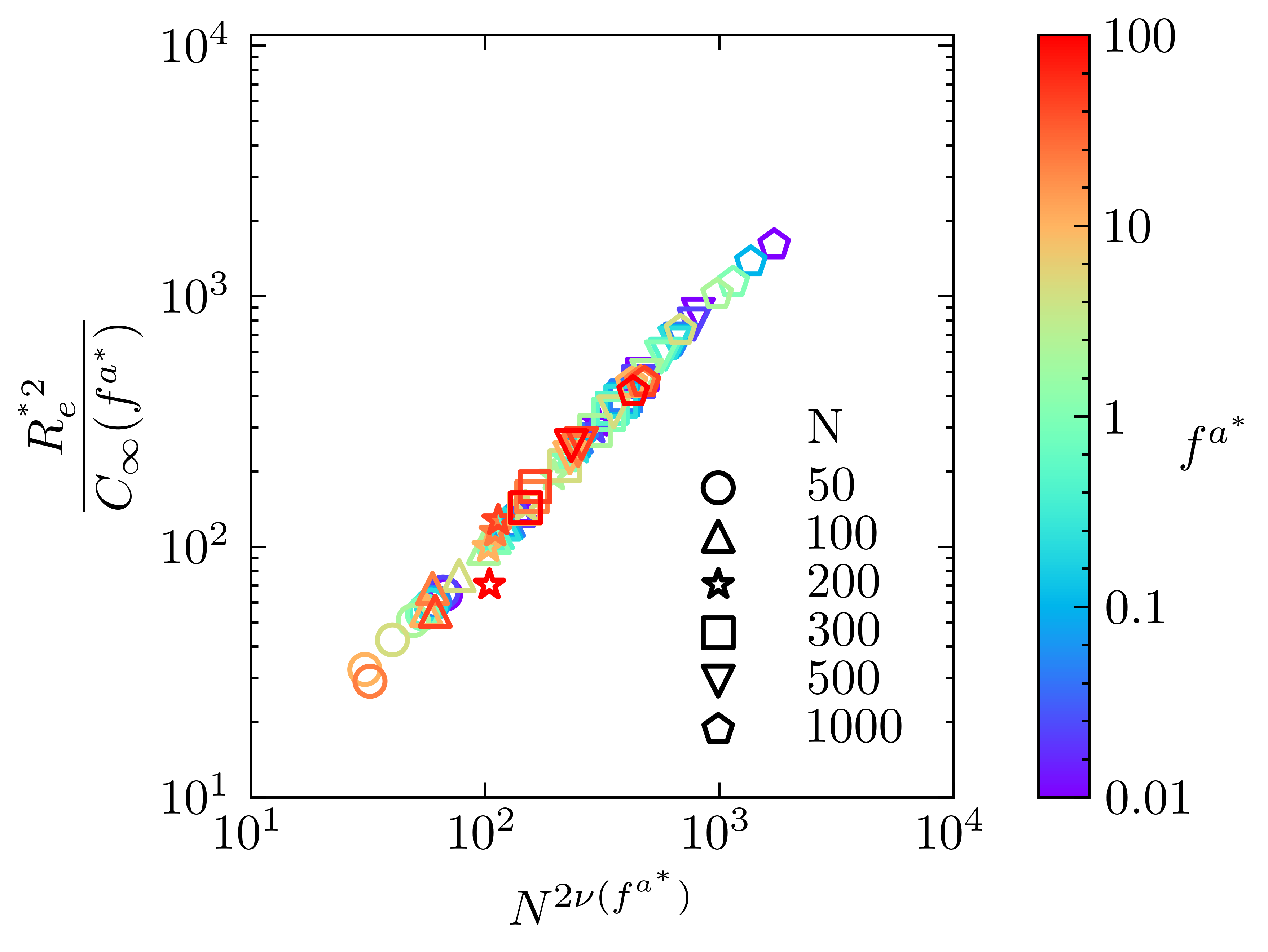}
    \caption{The normalised $\dfrac{R_{e}^{ ^{*}2}}{ C_{\infty}(f^{a^{*}})}$ as a function of $N^{2\nu (f^{a^{*}})}$ for underdamped chains. We excluded the outliers of $R_{e}^{ ^{*}}$, which are $(N,f^{a^{*}})\in \{ (50,50),(50,100),(100,100) \}$.}
    \label{fig:ReCinfNu}
\end{figure} 
  
  \begin{figure}[h]
    \centering
    \begin{tikzpicture}
            \draw (0,0) node[inner sep=0]{\includegraphics[width=\linewidth ]{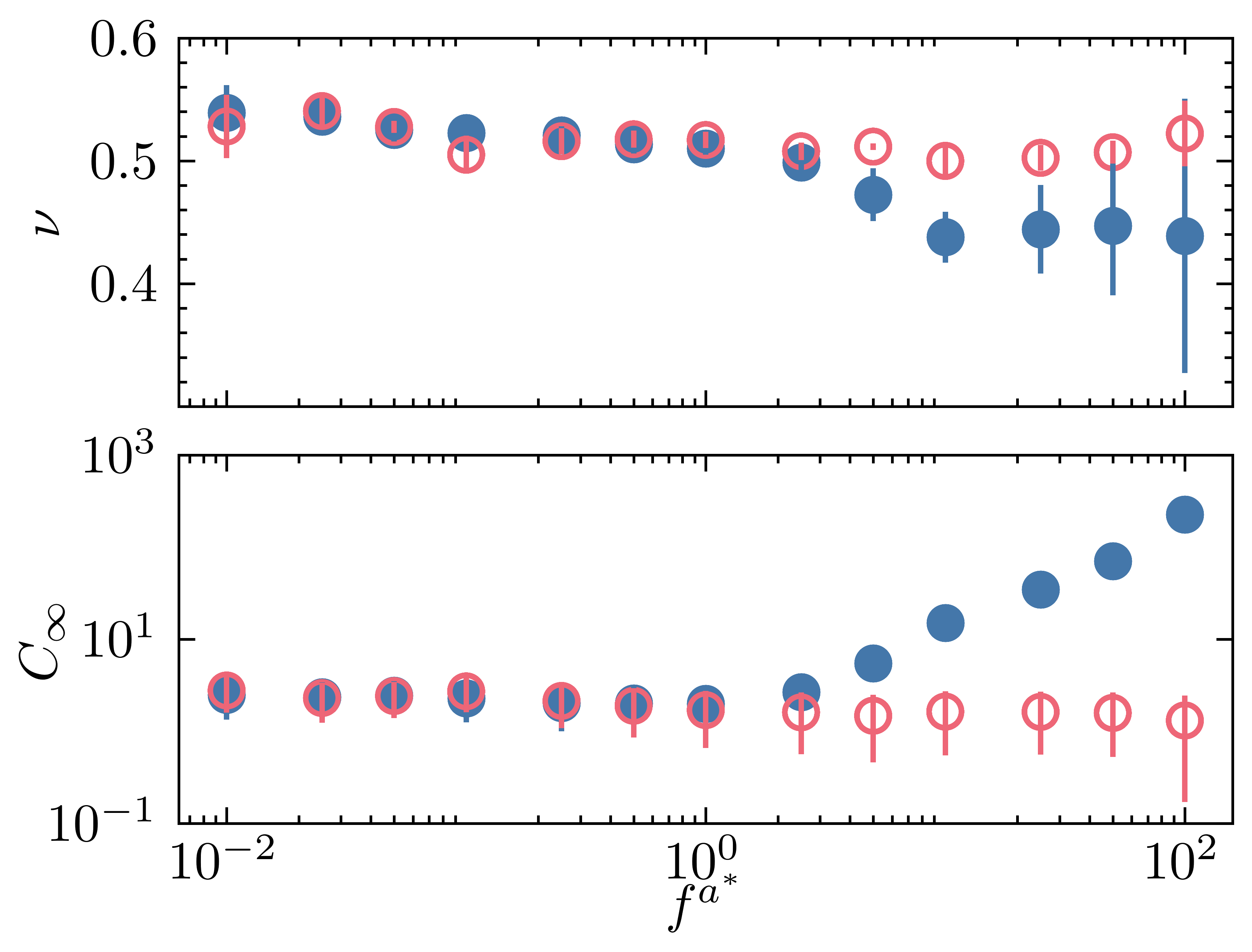}};
            \draw (3.3,2.74) node {\textbf{(a)}};
            \draw (3.3,-0.15) node {\textbf{(b)}};
            
    \end{tikzpicture}
    \vspace{-7mm}
    \caption{(a) Scaling exponent $\nu$ of end-to-end distance  and (b) generalized Flory constant $C_{\infty}$ as a function of $f^{a^{*}}$, assuming  $R_e^{ ^{*}2}(N,f^{a^{*}})= C_{\infty}(f^{a^{*}}) N^{2\nu(f^{a^{*}})}$.  The bold and empty symbols correspond to underdamped and overdamped chains, respectively. } 
    \label{fig:NuCinf}
    \vspace{-5mm}
\end{figure}

To investigate the scaling behavior of active polymers, we analyze  the dependence of  end-to-end distance on $N$ for different $f^{a^{*}} $. Assuming a scaling ansatz $\langle \vec{R}_e^{*2} \rangle=C_{\infty}(f^{a^{*}} ) N^{2\nu(f^{a^{*}} )}$, all the data for $N \ge 100$ can be collapsed onto a single master curve; see Fig.~\ref{fig:ReCinfNu}.  The dependence of scaling exponent $\nu$ on $f^{a^{*}} $ is shown in Fig.~\ref{fig:NuCinf}(a).  For $f^{a^{*}}<1$, $\nu$  
drops from  $\nu(f^{a^{*}}=0.01)=0.528$ close to that of the self-avoiding walk $\nu_{SAW}= 0.588$   to the ideal chain exponent $\nu_{ideal}=1/2$ for both overdamped and underdamped dynamics.  At larger activities, when $\tau_m^{*}/ \tau_{adv}^{*}=f^{a^{*}} >1$ the scaling exponent of overdamped active polymers remains constant within the error bars,  whereas  for inertial  chains the scaling exponent keeps on decreasing even further until $f^{a^{*}}=10$, and remains constant afterwards, $\nu_{inertial}(f^{a^{*}} \gg 1)\approx 0.44$.

 The $C_{\infty}(f^{a^{*}} )$ in the scaling ansatz is a generalized activity-dependent Flory constant which accounts for conformational rigidity of  polymer backbone.  Fig.~\ref{fig:NuCinf}(b) shows the extracted $C_{\infty} $ as a function of $f^{a^{*}}$ for active polymers with underdamped and overdamped dynamics. $C_{\infty}\sim 1$ up to $f^{a^{*}} =1$, whereas for $f^{a^{*}}>1$ it remains constant in the case of overdamped polymers, however it  rises steeply for inertial ones. A Flory constant $C_{\infty} > 1$ implies an increased conformational rigidity compatible with more straight conformations found for inertial active polymers, see the case of $f^{a^{*}}=100$ in Fig.~\ref{fig:conformation}(a).  
 A reduced $\nu$ combined with an increased $C_{\infty}(f^{a^{*}})$ at large activities  means that the average conformation of inertial active polymers resembles that of compact globule-like polymers with a large persistence length. 

\begin{figure*}[t]
    \centering
    \begin{tikzpicture}
       \draw (0,0) node[inner sep=0]{
      \includegraphics[width=0.9\linewidth]{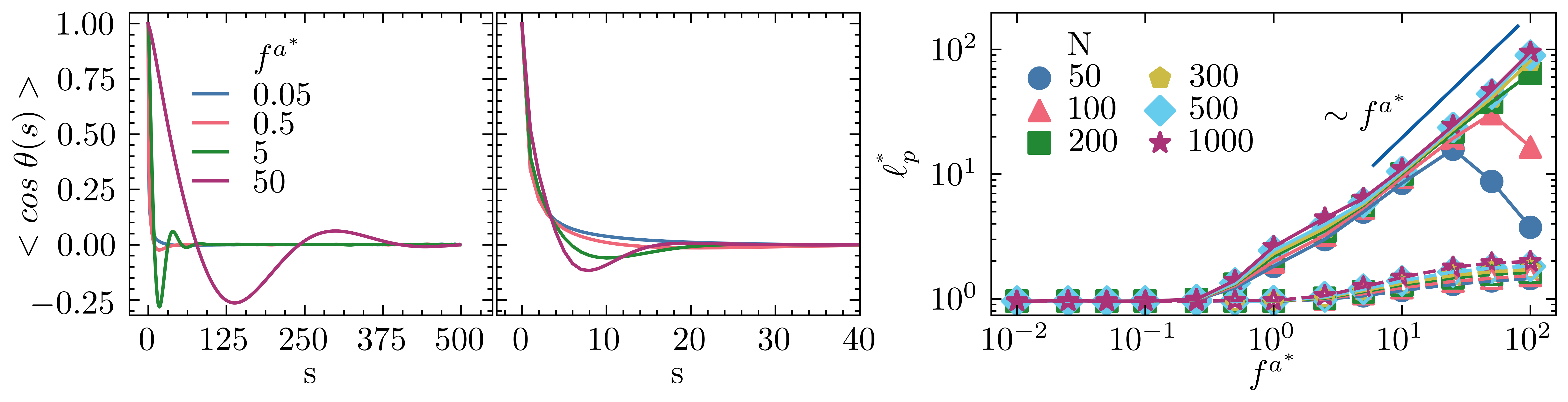} };
       \draw (-3.8,1.7) node {\textbf{(a)}};
           \draw (0.2,1.7) node {\textbf{(b)}};
           \draw (7.1,1.7) node {\textbf{(c)}};
        \end{tikzpicture}

    \caption{(a) and (b) Bond-bond orientational correlation function against curvilinear distance $s$ for inertial and overdamped chains with  $N=500$ at different activities, respectively. 
    (c) Dependence of persistence length  $\ell_p^*$ on $f^{a^{*}}$  for inertial (bold symbols) and overdamped (empty symbols) polymers of different  lengths.  
    }
    \label{fig:bond_corr}
    
\end{figure*}
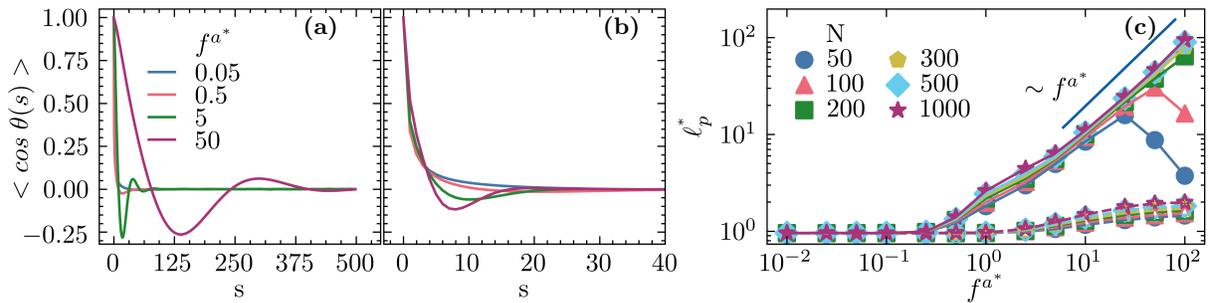

\begin{figure}[h!]
    \centering
    \centering
    \begin{tikzpicture}
            \draw (0,0) node[inner sep=0]{\includegraphics[width=\linewidth ]{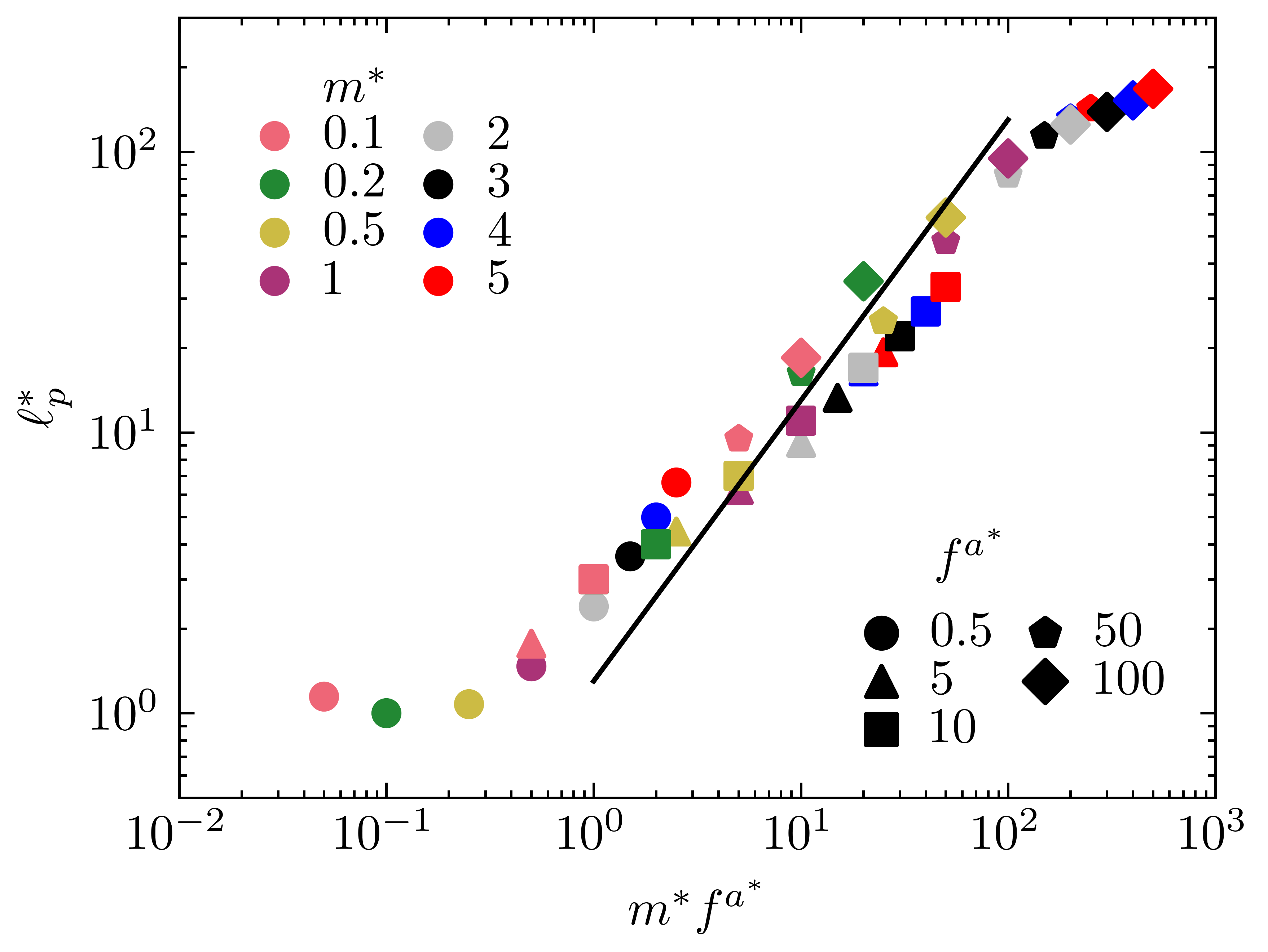}};
            \draw (0,-6.3) node {\includegraphics[width=0.95\linewidth]{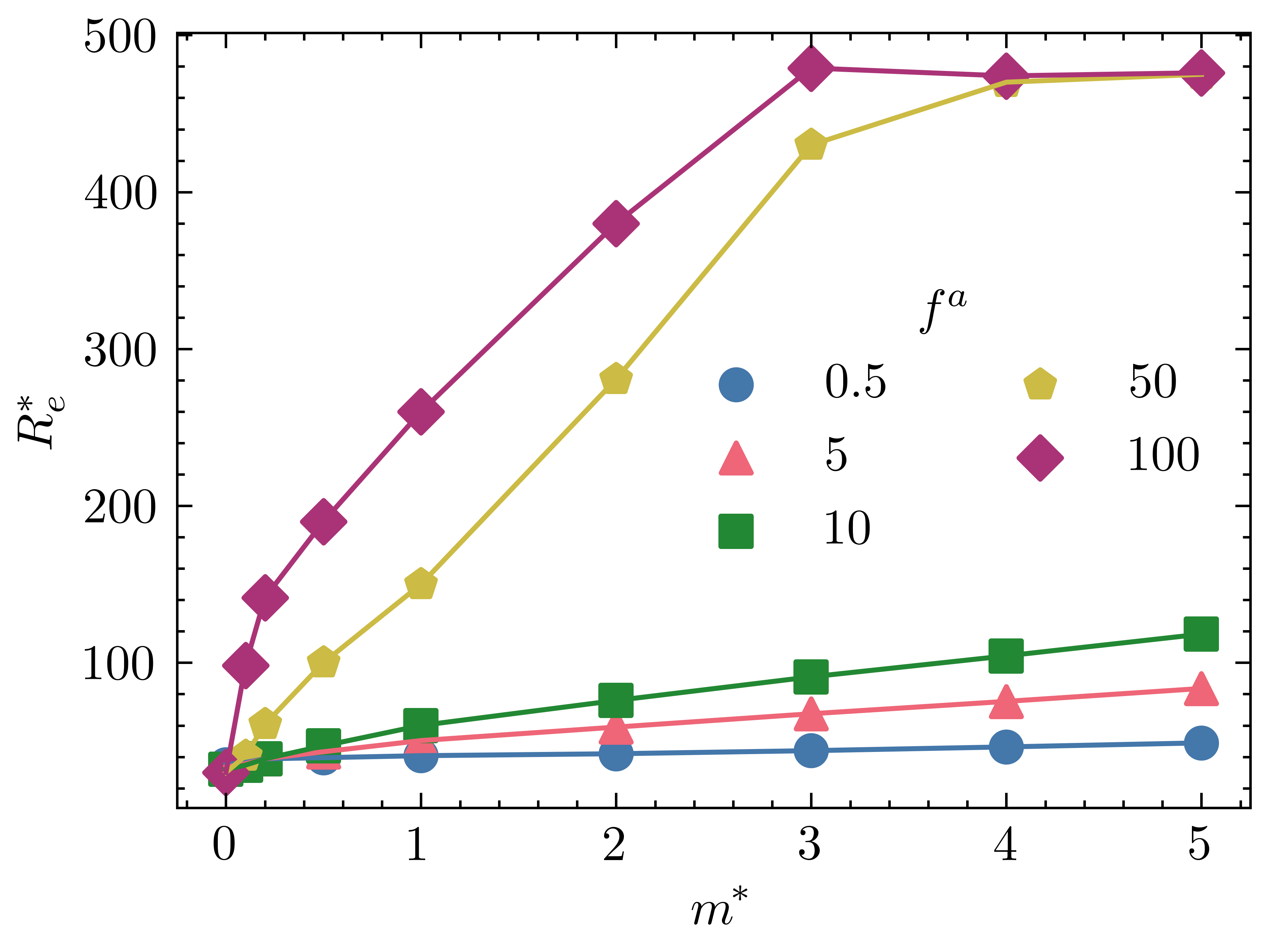}};
            \draw (3.5,1.75) node {\textbf{(a)}};
            \draw (3.5,-4.6) node {\textbf{(b)}};
    \end{tikzpicture}
    \caption{(a) The persistence length of active chains of length $N=500$ with different masses and activities as a function of the their mass times their active force $m^{*}f^{a^{*}}$. The solid line shows a fit with slope of 1 for the data between $1 \le m^{*}f^{a^{*}} \le 100$. (b) The end-to-end distance against mass for chains with $N=500$ and various activities.}
    \label{fig:lpMreM}

\end{figure}

A  growing  $C_{\infty} $ also implies an increasing  persistence length  with $f^{a^{*}}$. To obtain the persistence length, we compute the orientational bond-bond correlation function $\langle \cos \theta(s) \rangle$  defined as the cosine of the angle between any two bonds whose curvilinear distance is $s$.  Figs.~\ref{fig:bond_corr} (a) and ~\ref{fig:bond_corr}(b) present the $\langle \cos \theta(s) \rangle$ against curvilinear distance $s$ along the polymer backbone at different activities for inertial and overdamped  chains, respectively. For both cases, the bond-bond correlation function develops a negative dip at intermediate $s$, indicating a local  back-folding of active chains. For inertial chains, the bond orientational correlation  along backbone becomes larger upon increase of activity and  the negative dip appears at larger $s$,  whereas for overdamped active chains the decay length  is weakly affected by activity. $\langle \cos \theta(s) \rangle$ does not decay exponentially. Nonetheless, we choose to define persistence length $\ell_p^*$ as the curvilinear distance at which $\langle \cos \theta(\ell_p^*) \rangle=1/e$.  The extracted persistence lengths $\ell_p^*$s for overdamped and inertial chains versus $f^{a^{*}}$ are presented in  Fig.~\ref{fig:bond_corr} (c) and show identical trends for different $N$. For overdamped  polymers, $\ell_{p}^{ ^{*}}$ remains unity up to $f^{a^{*}} \sim 1$ and increases weakly beyond  it,  whereas for inertial  polymers $\ell_{p}^{ ^{*}}$ increases steeply with  $f^{a^{*}}$ when  $\tau_m^*/ \tau_{adv}^* >1$.

The inertia-induced persistence length at large activities scales linearly with $f^{a^{*}}$ and it can be interpreted as the distance at which an active monomer travels by speed $v^{a^{*}}=f^{a^{*}}/\gamma^*$ during inertial timescale $\tau_m^* $ giving rise to $\ell_{p} \sim f^{a^{*}} m^*/ {\gamma^*}^2 $ yielding  $\ell_{p}^{*} \sim f^{a^{*}} m^* $.  For $N=50$ and 100,  $\ell_{p}^{*}$ exhibits a decrease  when the contour length becomes comparable to the persistence length $ \ell_{p}^{*}\sim N$ and it cannot grow anymore, hence a new instability appears. Overall, we recognize a similar trend for  $\ell_{p}^{ ^{*}}$ and $C_{\infty}$ against active force corroborating the surmise of an inertia-induced bending rigidity in active polymers. We emphasize that locally straight conformations arise due to interplay between inertia and activity. Neither tangential driving force nor inertia alone does not lead to locally straight chain conformations as evidenced by $\ell_{p}^{ ^{*}}$ plot in Fig.~\ref{fig:bond_corr} (c).  
  To verify our suggested scaling for the activity-dependent persistence length $\ell_{p}^{*} \sim f^{a^{*}} m^* $ at large activities, we  vary both mass and active force for a fixed chain length $N=500$. Fig.~\ref{fig:lpMreM}(a) shows the persistence length as a function of $m^{*}f^{a^{*}}$ and confirms that   $\ell_{p}^*$ varies linearly with $m^{*}f^{a^{*}}$   when $1 \le m^{*}f^{a^{*}} \le 100$.
     For smaller values of $m^{*}f^{a^{*}}$ the relation is non-linear because $\tau_{m}^{*} < \tau_{adv}^{*}$ and the inertial effects are not dominant. On the other extreme of  large  $m^{*}f^{a^{*}}$, we observe deviation from linear scaling   because the magnitude of the persistence length becomes comparable to the chain length $N=500$. At this range of activity and mass the end-to-end distance reaches its largest possible value ($R_{e}^{ ^{*}}=N=500$) as shown in Fig.~\ref{fig:lpMreM}(b)  for the chains with $f^{a^{*}}\geq 50$ and $m^*\geq3$.

\subsection{Inertial effects on dynamics}
\begin{figure}[t]
    \centering
   \centering
    \begin{center}
        \begin{tikzpicture}
            \draw (0,0) node[inner sep=0]{\includegraphics[width=1\linewidth]{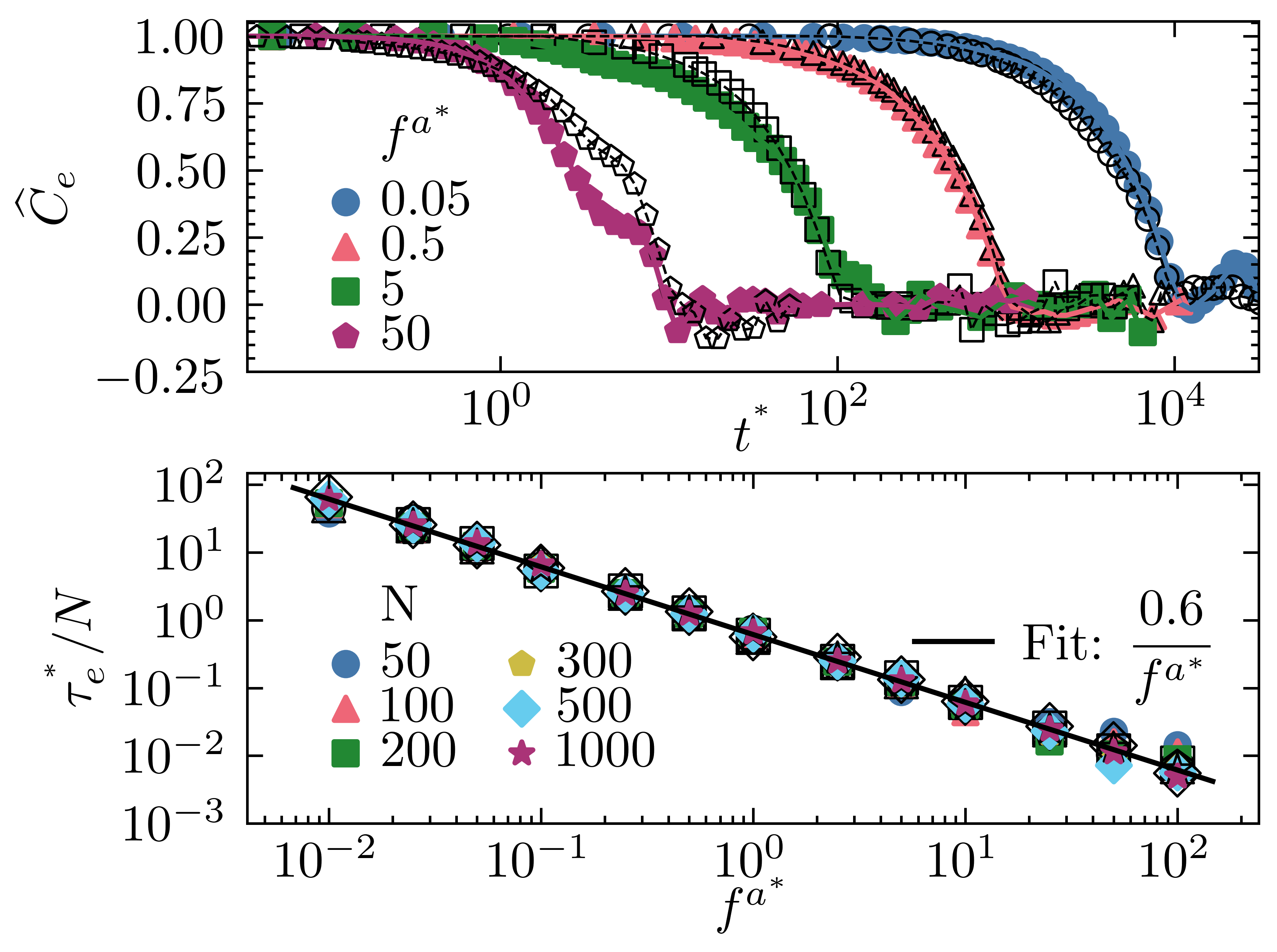}};
            \draw (3.7,2.8) node {\textbf{(a)}};
            \draw (3.7,-.4) node {\textbf{(b)}};
        \end{tikzpicture}
    \end{center}

    \caption{(a) Normalized end-to-end vector TACF, $ \widehat{C}_{e}$ as function of time $t$, for inertial (bold symbols) and overdamped  (empty black symbols) chains of  $N=500$ at different activities. (b) Corresponding  relaxation times divided by chain length against active force for inertial (bold symbols) and overdamped (empty black symbols), extracted from fitting the initial decay of $ \widehat{C}_{e}(t)$ with the exponential function.}
        \label{fig:dynamic_corr}

        \end{figure}

Having explored the inertial effects on structural features of active polymers,  we examine signatures of inertia on the dynamics. We start by  investigating the polymer orientational dynamics. To this end, we compute the time auto-correlation function (TACF) of the end-to-end vector normalized by its mean-squared value  in the steady-state limit $ \widehat{C}_{e}(t)=\lim_{t'\to \infty} \langle \vec{R}_{e}(t+t') \cdot \vec{R}_{e}(t') \rangle \big /\langle \vec{R}_e^2(t')\rangle $. As the  total active force $\vec{F}^{a}$ is proportional to the end-to-end vector, $\vec{F}^{a}=\sum_{i=1}^{N} \vec{f}^a_i=\frac{f^a}{\ell} \vec{R}_{e} $, the relaxation time of $\widehat{C}_{e}$ is identical to persistence time of  total active force.
 Fig.~\ref{fig:dynamic_corr}(a) shows $\widehat{C}_{e}$ at different activities for inertial and overdamped chains of length $N=500$. Although for large $f^{a^{*}}$, 
 the $ \widehat{C}_{e}(t)$ of inertial chains exhibits a decay with  oscillatory behavior,  the overall decorrelation timescales of  $\widehat{C}_{e}(t)$ of the two kinds of dynamics are very similar in contrast to huge difference in their chain conformations. 
This trend suggests that we have a universal relaxation time $\tau_e(N,f^a)$, which only depends on the strength of active force and the chain length and is independent of the mass. We find that the initial decay of  $\widehat{C}_{e}(t)$ for both overdamped and underdamped polymers can be approximately described by an exponential function, see Fig.~S2 in~\cite{SI}.  The extracted relaxation times $\tau_e^*$,  shown in Fig.~\ref{fig:dynamic_corr}(b), follow the scaling relation  
\vspace{-1.5mm}
\begin{equation}
\tau_e^* (N, f^{a^{*}}) \approx 0.6 N/f^{a^{*}},
\label{eq:tau_e}
\vspace{-1.5mm}
\end{equation}
 verifying that activity reduces relaxation time as $1/f^{a^{*}}$ and also  dependence of relaxation time on chain length becomes linear which is weaker than $N^{1+2\nu}$ dependence for self-avoiding Rouse model~\cite{Barkema2014}. These findings  are in agreement with reports of reference~\cite{Active_flexible} for overdamped active polymers.
  
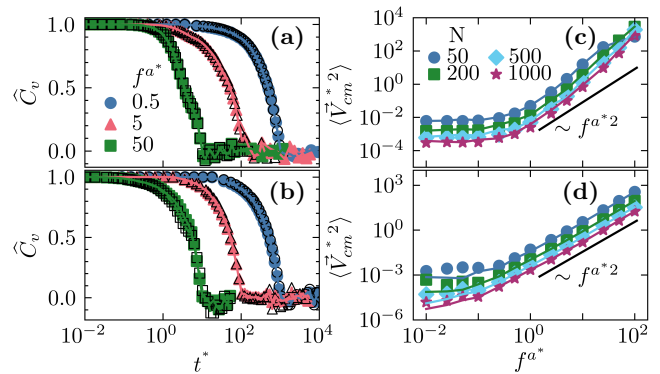
\begin{figure}[t]
    \centering
    \begin{center}
        \begin{tikzpicture}
            \draw (0,0) node[inner sep=0]{\includegraphics[width=1\linewidth]{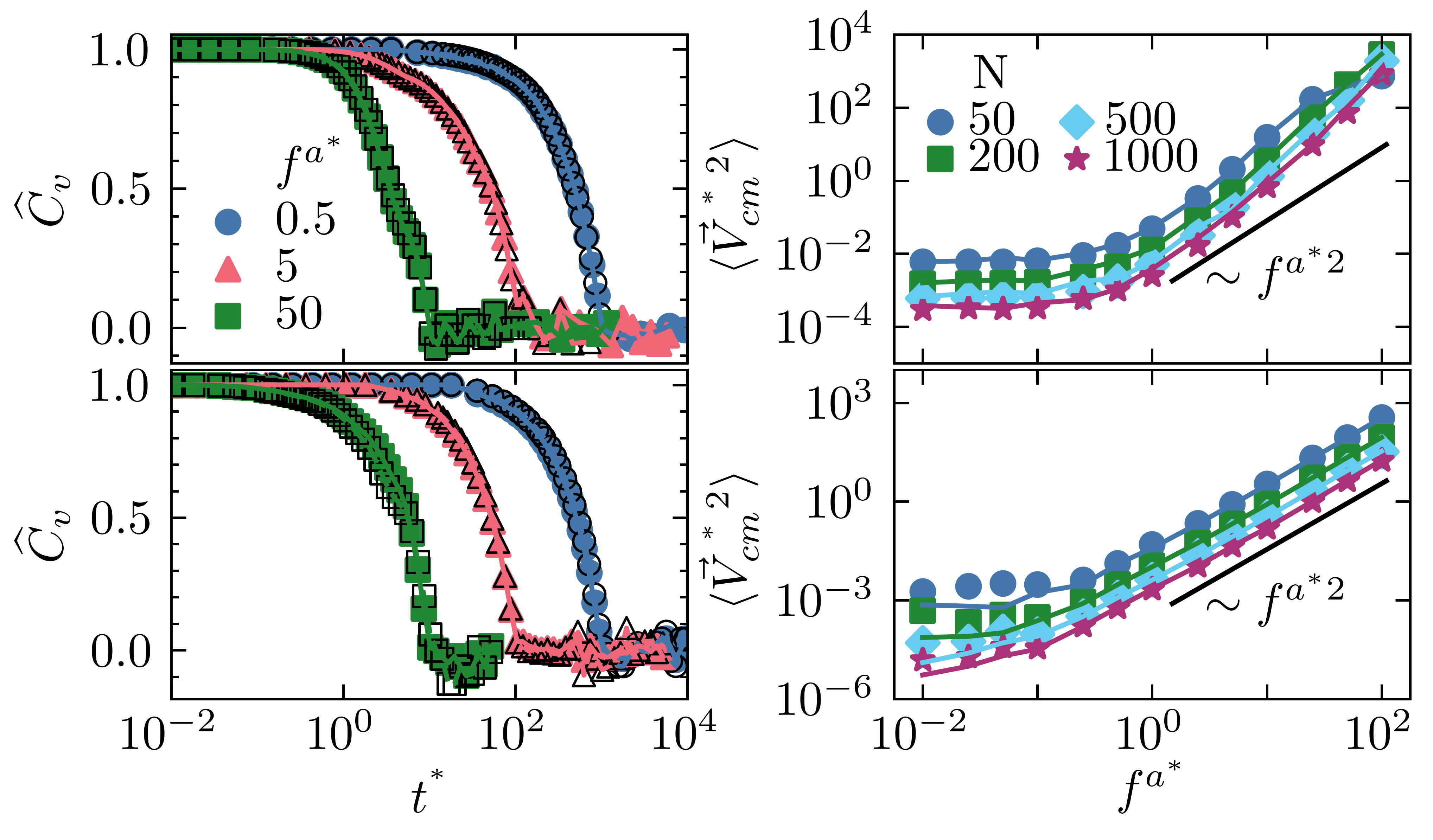}};
            \draw (-0.6,2.05) node {\textbf{(a)}};
            \draw (3.3,2.05) node {\textbf{(c)}};
            \draw (-0.6,0) node {\textbf{(b)}};
            \draw (3.3,0) node {\textbf{(d)}};
        \end{tikzpicture}
    \end{center}
    \caption{(a) and (b) Normalized Center of mass velocity TACF for inertial and  overdamped active polymers of  $N=500$ and different activities, respectively. The filled symbols are obtained directly from simulations, whereas  empty black ones are calculated using Eq.~\eqref{eq:vcorrFinal} and Eq.~\eqref{eq:vcorr:BD} with $C_e$ as an input from simulations.
  (c) and (d) The  mean-squared velocity of center of mass of different chain lengths as given in the legends versus $f^{a ^{*}}$ for inertial and overdamped active polymers, respectively. The lines show theoretical predictions of Eq.~\eqref{eq:V_cm2} and Eq.~\eqref{eq:Vcm2BD}. }
        \label{fig:vel_corr}
\end{figure}

We subsequently compute TACF of the center of mass velocity normalized by the  long-time mean-squared velocity of center of mass in the steady-state limit $\widehat{C}_{v}(t)=\lim_{t'\to \infty} \langle \vec{V}_{\text{cm}}(t+t') \cdot \vec{V}_{\text{cm}}(t') \rangle \big / \langle \vec{V}_{\text{cm}}^2(t') \rangle $, where  $\vec{V}_{\text{cm}}(t)=\frac{1}{N}\sum_{i=1}^{N} \vec{v}_i$. Fig.~\ref{fig:vel_corr}(a) and (b) show $\widehat{C}_{v}$ for underdamped and overdamped active polymers of $N=500$ at different $f^{a^{*}}$, respectively. Unlike passive systems, we observe a finite relaxation of  $\widehat{C}_{v}$ even for the overdamped dynamics  manifesting the contribution of active forces on velocity correlations. For the underdamped active chains, we recognize an initial fast decay with a relaxation time  $\tau_m^*=1$ in addition to a longer decay time which is similar to that of the overdamped chains.  For both kinds of dynamics, the characteristic relaxation times of $\widehat{C}_{v}$   decrease with $f^{a^{*}}$, similar to trends observed for $\widehat{C}_{e}$. 
In Figs.~\ref{fig:vel_corr} (c) and (d), we have presented   $\langle \vec{V}_{\text{cm}}^{*2} \rangle$  as a function of $f^{a^{*}}$ for underdamped and overdamped dynamics, respectively. For inertial chains,  the mean-squared velocity is almost independent of $f^{a^{*}}$ for  $f^{a^{*}}<1$, whereas for large active forces  it grows steeper than $f^{a^{*}2}$.  In the overdamped scenario, we observe a  weak decrease of $\langle \vec{V}_{\text{cm}}^{*2} \rangle$ for small $f^{a^{*}}$ but it grows as $f^{a^{*}2}$ for larger activities. Nevertheless, the mean-squared velocities of inertial chains at large activities are one order of magnitude larger. 

\begin{figure}[t]
    \centering
    \centering
    \begin{tikzpicture}
            \draw (0,0) node[inner sep=0]{\includegraphics[width=\linewidth ]{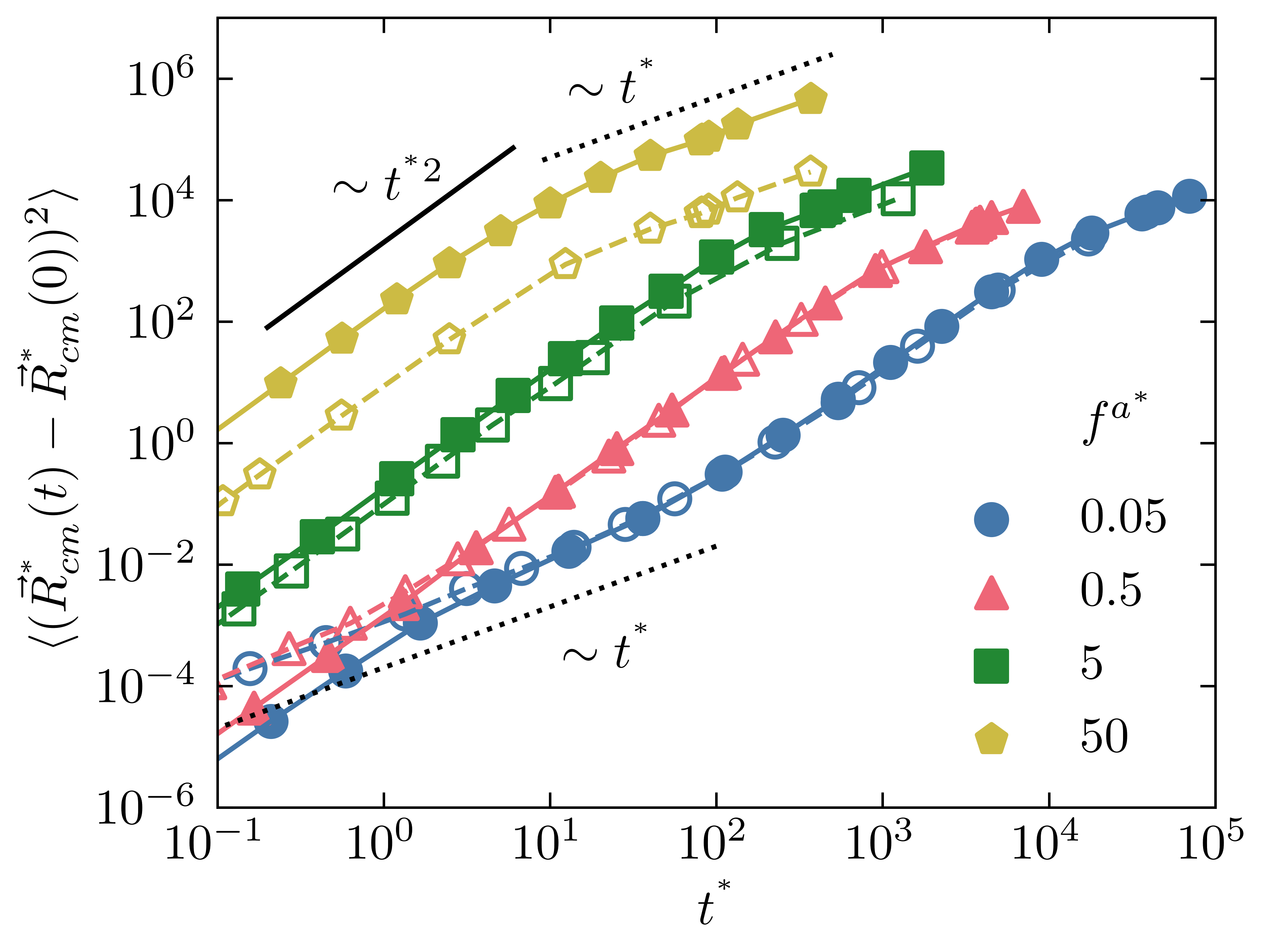}};
            \draw (0,-6.3) node {  \includegraphics[width=0.95\linewidth]{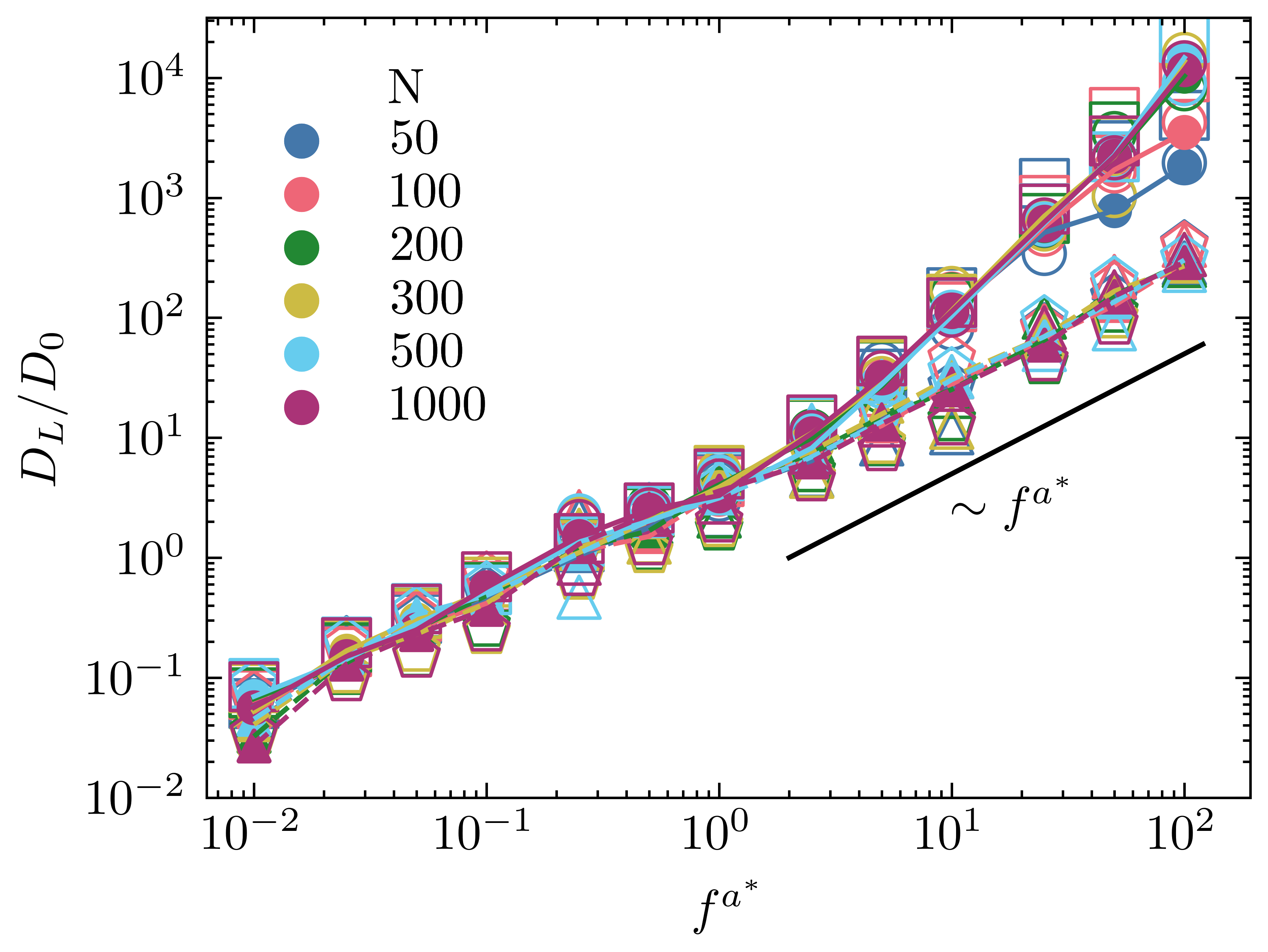}};
            \draw (3.,2.75) node {\textbf{(a)}};
            \draw (3.,-3.6) node {\textbf{(b)}};
    \end{tikzpicture}
    \caption{(a) Mean squared displacement of center of mass for chains of $N=500$ and different activities for underdamped (filled) and overdamped (empty) polymers. (b) Long-time diffusion coefficient $D_L$ of the center of mass normalized by monomer diffusion $D_0$ against $f^{a^{*}}$ for active polymers of chain lengths $50 \le N \le 1000$ in the underdamped (bold circles) and overdamped (bold triangles) regimes.
   The empty circles and triangles show predictions of Eq.~\eqref{eq:Dlong_LD} and Eq.~\eqref{eq:Dlong_BD}, receptively, using   end-to-end vector TACF  from simulations. The   squares and pentagons display the prediction of 
   Eq.~\eqref{eq:Dexplicit}, assuming $\tau_e^*=0.6N/f^{a^{*}}$.}
 
    \label{fig:msd-DL}
\end{figure}

Finally, we investigate the mean-squared displacement (MSD) of the center of mass as presented in Fig.~\ref{fig:msd-DL} (a) for the active chains of length $N=500$ with   overdamped and underdamped dynamics at different activities. For small activities   $f^{a^{*}}< 1$, the MSDs of the two scenarios at times $t^*>1$ coincide.     After an initial short-time diffusive (overdamped) or ballistic (underdamped) regime, the MSDs exhibit a superdiffusive regime  induced by activity, followed by a long-time diffusion in agreement with the results reported in~\cite{Active_flexible} for overdamped active polymers. However, at larger activities $f^{a^{*}}\geq 5$, the MSDs of two scenarios start to depart and we observe a  remarkably faster dynamics for underdamped chains. At large activities the MSDs of both underdamped and overdamped chains exhibit a crossover from  ballistic regime directly to diffusive regime.
We extract the long-time diffusion coefficient of center of mass $D_L$ from the slope of mean-squared displacement at large times, {\it i.e.}, $t^* > \tau_e^* $.  Fig.~\ref{fig:msd-DL}(b) shows $D_L$ versus $f^{a ^{*}}$ for inertial and overdamped chains.  Intriguingly,  two distinct sets of curves for inertial and overdamped chains emerge and the long-time diffusion of inertial chains at large activities is two orders of magnitude larger. Notably, $D_L$ is almost  independent of chain length,  in contrast to passive polymers for which $D_L \sim 1/N$~\cite{polymerDoi},  depending mainly on $f^{a^{*}}$ and \emph{mass}. Our results in the overdamped limit are consistent with an earlier reports of reference~\cite{Active_flexible} 
but clearly show that inertia leads to additional enhancement of the long-time dynamics of active polymers and raise questions about its origin. To rationalize the remarkable enhancement of dynamics, we put forward analytical calculations which illuminate the relationship between TACFs of the center of mass velocity and the total active force $\vec{F}^a$ in the steady-state limit for any active polymer model as presented in the following subsection. Then, we will  compare theoretical predictions with our simulation results for the tangentially-driven polymers.

  \subsection{Analytical calculations of center of mass dynamics}   
  The dynamics of center of mass velocity  $\vec{V}_{\text{cm}}(t)$  is obtained by summing the equations of motions of  all the beads given by Eq.~\eqref{eq:Langevin}. 
As contributions from the internal forces cancel out,  the  equation  for $\vec{V}_{\text{cm}}(t)$ simplifies to:
\begin{equation}
  m\dot{\vec{V}}_{\text{cm}}=  -\gamma \vec{V}_{\text{cm}}+\frac{1}{N} (\vec{F} ^a+\vec{F}^{r} ),
\end{equation}
in which $\vec{F} ^r=\sum_{i=1}^N \vec{f}_i^r$ is a sum of all the random forces with a zero mean and $\langle \vec{F}^r(t) \cdot \vec{F}^r (t') \rangle=6 N D_0 \gamma^2 \delta(t-t')$.  
Integrating this  first-order differential equation yields
\begin{eqnarray}
   \vec{V}_{\text{cm}}(t) &=&\vec{V}(t_0) e^{ -\frac{\gamma}{m} (t-t_0)}\\ \nonumber
   &+&\frac{1}{N m}\int_{t_0}^t d\tau [\vec{F} ^a(\tau)+\vec{F}^{r}(\tau)] e^{ -\frac{\gamma}{m} (t-t_0-\tau)}. 
\end{eqnarray}
Using this solution, we calculate  the velocity TACF in the steady-state limit, defined as   $C_{v}(t)=\lim_{t'\to \infty} \langle \vec{V}_{\text{cm}}(t+t') \cdot \vec{V}_{\text{cm}}(t') \rangle$, yielding
\begin{align}
    &C_{v}(t) = 
       \frac{3 \gamma D_0}{N m } \: e^{ \dfrac{-t}{\tau_m}} \nonumber\\
     & + \frac{1}{2 m \gamma N^2}       \int_{0}^{+\infty} \Big[\; C_{f}(u-t) \;+\;
      C_{f}(u+t)\; \Big] \;e^{\dfrac{-u}{\tau_m}} \; du,
   \label{eq:vcorrFinal}
   \end{align}
  in which $C_f(t)= \lim_{t' \to \infty}\langle \vec{F}^a(t+t')\cdot \vec{F}^a(t')  \rangle $ is the TACF of the total active force. The first term in Eq.~\eqref{eq:vcorrFinal} represents the passive diffusive contribution from random forces, which decays exponentially with the relaxation time $\tau_m$~\cite{inertia_Rouse}, whereas the second term entangles the correlation of the total active force with the inertial relaxation. This term reflects  \emph{memory effects} induced by active forces and accounts for the emergent inertial effects. In the limit of vanishing mass, $C_{v}$ simplifies to 
 \begin{equation}
      \lim_{m\rightarrow 0}  C_{v}(t>0) = (\frac{1}{N \gamma })^2 \;C_{f}(t).
       \label{eq:vcorr:BD}
 \end{equation}
 Concerning tangentially-driven polymers, for which the total active force $\vec{F} ^a=f^a\; \vec{R}_e/\ell$ is directly coupled to the chain conformation,  the total force correlation is $C_f(t)=({f^a}^2 \langle {R_e}^2\rangle /\ell^2)\,   \widehat{C}_e(t)$.
 Predictions of Eq.~\eqref{eq:vcorrFinal} and \eqref{eq:vcorr:BD}  when normalized by their values at $t^*=0$  show  excellent agreement with  velocity correlations computed from simulations as demonstrated in Fig.~\ref{fig:vel_corr}(a) and (b).

For an active force TACF with exponential decay $C_{f}=A_f e^{-t/\tau_f}$, Eq.~\eqref{eq:vcorrFinal} simplifies to
\vspace{-2.5mm}
 \begin{align}
   C_{v}(t)&= \frac{3 \gamma D_0}{N m } \: e^{ \dfrac{-t}{\tau_m}}  \label{eq:vcorr_exp}  \\
   &+ \frac{A_f^{2}}{(N m)^2}  
       (\dfrac{\tau_m^{2}\tau_f}{\tau_f^2 - \tau_m^2})\Big[ \tau_f\; e^{-t/\tau_f} - \tau_m\; e^{-t/\tau_m}  \Big], \nonumber
       \vspace{-2.5mm}
 \end{align}
 which consists of  exponential decays at two timescales; the inertial time $\tau_m$ and   decay time of active force TACF $\tau_f$. In this case, the center of mass velocity TACF of active polymers becomes identical to that of inertial active Ornstein–Uhlenbeck model~\cite{Inertial_AOUP,lowen-AOUP}. The approximation of exponential decay of active force TACF becomes exact in the case of an active Brownian polymer model~\cite{ABPO_ANALYTICS-eisenstecken_gompper_winkler_2016,ABPO-N200} with only translational inertia, see Sec.\,III in SM~\cite{SI}.  As discussed earlier, $\widehat{C}_e(t)$ approximately follows an exponential decay. Thus, Eq.~\eqref{eq:vcorr_exp} also provides  a good description of tangentially-driven active polymers provided that we consider an activity-dependent amplitude $A_f$ which is coupled to polymer mean conformation. \\
 We can also obtain the  steady-state mean-squared center of mass velocity for tangentially-driven polymers  from Eq.~\eqref{eq:vcorrFinal}, yielding 
 \begin{align}
    &\lim_{t \to \infty} \langle \vec{V}_{\text{cm}}^2(t) \rangle=\nonumber \\ &\frac{3 \gamma D_0}{N m } + \frac{{f^a}^2}{m \gamma N^2} \frac{\langle{R_e}^2\rangle }{ \ell^2} 
        \int_{0}^{+\infty}  \widehat{C}_{e}(u)  \;e^{\dfrac{-u}{\tau_m}} \; du 
        \label{eq:V_cm2}
       \end{align}
    which for small $f^a$ is dominated by the diffusive term, whereas when $ \frac{{f^a}^2}{3 N \gamma^2 D_0} \frac{\langle{R_e}^2\rangle }{ \ell^2} >1  $, is determined by the active contribution. For sufficiently large $f^a$  such that the relaxation time $\tau_e < \tau_m$,
    $\int_{0}^{+\infty}  \widehat{C}_{e}(u)  \;e^{\frac{-u}{\tau_m}} du  \approx \tau_{e}$. In this limit, $ \langle \vec{V}_{\text{cm}}^2 \rangle (f^a \gg 1)= \frac{{f^a}^2}{m \gamma N^2} \frac{\langle{R_e}^2  \rangle} {\ell^2} \tau_{e}$.  Using  Eq.~\eqref{eq:tau_e}   and the scaling behavior of $\langle{R_e}^2\rangle$ it becomes proportional to ${f^a}^2 N^{2\nu(f^a)-1} $.
  Eq.~\ref{eq:V_cm2} in the limit of vanishing mass simplifies to
    \begin{equation}
        \lim_{m\rightarrow 0} \langle \vec{V}_{\text{cm}}^2(t \gg 1) \rangle= \frac{ f^{a2}\langle R_e^2 \rangle}{(\ell^2 N \gamma)^2 } .
        \label{eq:Vcm2BD}
    \end{equation}
    Comparison of Eq.\eqref{eq:V_cm2} and Eq.~\eqref{eq:Vcm2BD} with  $\langle \vec{V}_{\text{cm}}^ { ^{*}2}(0) \rangle$, presented in Fig. \ref{fig:vel_corr} (c) and (d), shows very good agreement.
  
The explicit form of the TACF of velocity given by Eq.~\eqref{eq:vcorrFinal}, allows us to predict the long-time diffusion of center of mass by integrating it over time: 
$D_L=\frac{1}{3} \int_0^{\infty} dt   C_{v}(t)$.
For inertial tangentially-driven polymers, this yields
\begin{align}
&D_L=  \frac{  D_0  }{N  }+\frac{f^{a2}\langle R_e^2\rangle}{6 m \gamma N^2\ell^2} \label{eq:Dlong_LD}\\ 
&\times\;\;
        \int_{0}^{+\infty}dt  \int_{0}^{+\infty} \Big[ \widehat{C}_{e}(u-t) +
      \widehat{C}_{e}(u+t) \Big] e^{\dfrac{-u}{\tau_m}} du.  \nonumber
\end{align}
For the overdamped chains, in the limit of $m \to 0$, Eq. \eqref{eq:Dlong_LD} simplifies to
\begin{equation}
  D_L= \frac{  D_0  }{N  }+ \frac{f^{a2}\langle R_e^2\rangle}{ 3\gamma^2 N^2\ell^2} \int_{0}^{+\infty} dt \;\widehat{C}_{e}(t),
   \label{eq:Dlong_BD}
\end{equation}
yielding identical  results to calculations of reference~\cite{Active_flexible}.
Predictions of Eqs.~\eqref{eq:Dlong_LD} and ~\eqref{eq:Dlong_BD} are shown in Fig.~\ref{fig:msd-DL}(b)  by empty circles and triangles, respectively, demonstrating  excellent agreement with simulation results. Approximating $\widehat{C}_{e}(t)$ with an exponential function $\widehat{C}_{e}(t)=e^{-t/\tau_{e}}$ and using  Eq.~\eqref{eq:tau_e} for the relaxation time, Eqs.~\eqref{eq:Dlong_LD} and \eqref{eq:Dlong_BD}  yield  an identical approximate equation for inertial and overdamped systems: 
\begin{equation} 
D_L \approx \frac{  D_0  }{N  } +
       \frac{{f^a}^2 \langle R_e^2\rangle \tau_{e}}{3  N^2 \gamma^{2}\ell^2} .
   \label{eq:Dexplicit}
\end{equation}
 This equation clarifies that the inertia-induced enhancement of $D_L$  originates from extended chain conformations generating an overall larger $ \langle R_e^2\rangle $.
Assuming $\tau_e \sim N/f^a$, the long time diffusion scales as $D_L \sim N^{2 \nu-1}$ which weakly depends on $N$. Hence, this equation shows that at sufficiently large activities  the dominant contribution in $D_L$ comes from activity.  We note that in the overdamped limit, $\langle R_e^2\rangle$  becomes independent of activity at large active forces, see Fig.~\ref{fig:conformation}(b) and given that  $\nu \approx 0.5$,  $D_L$ becomes independent of $N$ and varies linearly with $f^a$ in agreement with prior simulation results of a similar model~\cite{Active_flexible} and 
analytical calculations for Gaussian tangentially-driven polymers~\cite{BASKARAN,Phillipps_22}. 
Comparison of  simulation results with Eq.~\eqref{eq:Dexplicit} are also included in Fig.~\ref{fig:msd-DL}(b),  revealing  good agreement and verifying the approximate relation between $D_L$, $\langle R_e^2\rangle$ and $\tau_{e}$.
\\\\
{\section{ Conclusion}
We have combined bead-spring simulations and analytical theory to provide new insights into the fundamental interplay between inertia, activity and flexibility in tangentially-driven active filaments. For this model, active forces  are coupled to polymer conformations and inertia conspicuously affects both conformational and dynamical features. Inertial effects become significant,  when the inertial timescale of a monomer becomes longer than   the timescale of its advection by active force causing  a delay between end-to-end vector and  center of mass velocity. 
Inertial collisions of high-speed monomers cause extended conformations, hence increasing the persistence length of polymers.  The chain unwinding  in turn results in enhanced mean velocity and diffusion of the center of mass, as elucidated by our theoretical analysis.  The analytical calculations show that for tangentially-driven polymers the key quantity determining the enhanced dynamics in active polymers is the ratio of end-to-end distance to contour length.  

It is noteworthy  that our theoretical predictions are valid for any active polymer model and any degree of flexibility although for  simplicity  we focused on active polymers with a  persistence length comparable to the bead size.    
Our findings provide a guideline for design of active robotic worms, for which inertial effects are non-negligible. 
Moreover, the results of this study inspire investigation of a deeper link between memory effects and inertia in the collective dynamics of active filaments and more generally other active systems with internal degrees of freedom.  
We conclude our paper with a final remark. Given the significant effects of inertia on active polymer conformation and long-time dynamics, the ratio of mass to damping coefficient  in simulations of active polymers with inertial terms~\cite{Locatelli_21,Prathyusha} should be chosen carefully to ensure that they capture the correct inertial or overdamped regime. \\\\
\section{ Acknowledgments }
 We acknowledge A. Deblais, C. Coulais and D. Bonn for fruitful discussions. The computations were carried out on the Dutch National e-Infrastructure with the support of the SURF Cooperative. This work was part of the D-ITP consortium, a program of the Netherlands Organization for Scientific Research (NWO) that is funded by the Dutch Ministry of Education, Culture and Science (OCW).
 
 \bibliography{active_polymer.bib}

\begin{thebibliography}{48}%
\makeatletter
\providecommand \@ifxundefined [1]{%
 \@ifx{#1\undefined}
}%
\providecommand \@ifnum [1]{%
 \ifnum #1\expandafter \@firstoftwo
 \else \expandafter \@secondoftwo
 \fi
}%
\providecommand \@ifx [1]{%
 \ifx #1\expandafter \@firstoftwo
 \else \expandafter \@secondoftwo
 \fi
}%
\providecommand \natexlab [1]{#1}%
\providecommand \enquote  [1]{``#1''}%
\providecommand \bibnamefont  [1]{#1}%
\providecommand \bibfnamefont [1]{#1}%
\providecommand \citenamefont [1]{#1}%
\providecommand \href@noop [0]{\@secondoftwo}%
\providecommand \href [0]{\begingroup \@sanitize@url \@href}%
\providecommand \@href[1]{\@@startlink{#1}\@@href}%
\providecommand \@@href[1]{\endgroup#1\@@endlink}%
\providecommand \@sanitize@url [0]{\catcode `\\12\catcode `\$12\catcode
  `\&12\catcode `\#12\catcode `\^12\catcode `\_12\catcode `\%12\relax}%
\providecommand \@@startlink[1]{}%
\providecommand \@@endlink[0]{}%
\providecommand \url  [0]{\begingroup\@sanitize@url \@url }%
\providecommand \@url [1]{\endgroup\@href {#1}{\urlprefix }}%
\providecommand \urlprefix  [0]{URL }%
\providecommand \Eprint [0]{\href }%
\providecommand \doibase [0]{http://dx.doi.org/}%
\providecommand \selectlanguage [0]{\@gobble}%
\providecommand \bibinfo  [0]{\@secondoftwo}%
\providecommand \bibfield  [0]{\@secondoftwo}%
\providecommand \translation [1]{[#1]}%
\providecommand \BibitemOpen [0]{}%
\providecommand \bibitemStop [0]{}%
\providecommand \bibitemNoStop [0]{.\EOS\space}%
\providecommand \EOS [0]{\spacefactor3000\relax}%
\providecommand \BibitemShut  [1]{\csname bibitem#1\endcsname}%
\let\auto@bib@innerbib\@empty
\bibitem [{\citenamefont {Ramaswamy}(2010)}]{Ramaswamy2010}%
  \BibitemOpen
  \bibfield  {author} {\bibinfo {author} {\bibfnamefont {S.}~\bibnamefont
  {Ramaswamy}},\ }\href {\doibase 10.1146/annurev-conmatphys-070909-104101}
  {\bibfield  {journal} {\bibinfo  {journal} {Annual Review of Condensed Matter
  Physics}\ }\textbf {\bibinfo {volume} {1}},\ \bibinfo {pages} {323} (\bibinfo
  {year} {2010})}\BibitemShut {NoStop}%
\bibitem [{\citenamefont {Marchetti}\ \emph {et~al.}(2013)\citenamefont
  {Marchetti}, \citenamefont {Joanny}, \citenamefont {Ramaswamy}, \citenamefont
  {Liverpool}, \citenamefont {Prost}, \citenamefont {Rao},\ and\ \citenamefont
  {Simha}}]{Marchetti1}%
  \BibitemOpen
  \bibfield  {author} {\bibinfo {author} {\bibfnamefont {M.~C.}\ \bibnamefont
  {Marchetti}}, \bibinfo {author} {\bibfnamefont {J.~F.}\ \bibnamefont
  {Joanny}}, \bibinfo {author} {\bibfnamefont {S.}~\bibnamefont {Ramaswamy}},
  \bibinfo {author} {\bibfnamefont {T.~B.}\ \bibnamefont {Liverpool}}, \bibinfo
  {author} {\bibfnamefont {J.}~\bibnamefont {Prost}}, \bibinfo {author}
  {\bibfnamefont {M.}~\bibnamefont {Rao}}, \ and\ \bibinfo {author}
  {\bibfnamefont {R.~A.}\ \bibnamefont {Simha}},\ }\href {\doibase
  10.1103/RevModPhys.85.1143} {\bibfield  {journal} {\bibinfo  {journal} {Rev.
  Mod. Phys.}\ }\textbf {\bibinfo {volume} {85}},\ \bibinfo {pages} {1143}
  (\bibinfo {year} {2013})}\BibitemShut {NoStop}%
\bibitem [{\citenamefont {Romanczuk}\ \emph {et~al.}(2012)\citenamefont
  {Romanczuk}, \citenamefont {Bär}, \citenamefont {Ebeling}, \citenamefont
  {Lindner},\ and\ \citenamefont {Schimansky-Geier}}]{ABP-review}%
  \BibitemOpen
  \bibfield  {author} {\bibinfo {author} {\bibfnamefont {P.}~\bibnamefont
  {Romanczuk}}, \bibinfo {author} {\bibfnamefont {M.}~\bibnamefont {Bär}},
  \bibinfo {author} {\bibfnamefont {W.}~\bibnamefont {Ebeling}}, \bibinfo
  {author} {\bibfnamefont {B.}~\bibnamefont {Lindner}}, \ and\ \bibinfo
  {author} {\bibfnamefont {L.}~\bibnamefont {Schimansky-Geier}},\ }\href
  {\doibase 10.1140/epjst/e2012-01529-y} {\bibfield  {journal} {\bibinfo
  {journal} {The European Physical Journal Special Topics}\ }\textbf {\bibinfo
  {volume} {202}},\ \bibinfo {pages} {1} (\bibinfo {year} {2012})}\BibitemShut
  {NoStop}%
\bibitem [{\citenamefont {Scholz}\ \emph {et~al.}(2018)\citenamefont {Scholz},
  \citenamefont {Jahanshahi}, \citenamefont {Ldov},\ and\ \citenamefont
  {Löwen}}]{Scholz_18}%
  \BibitemOpen
  \bibfield  {author} {\bibinfo {author} {\bibfnamefont {C.}~\bibnamefont
  {Scholz}}, \bibinfo {author} {\bibfnamefont {S.}~\bibnamefont {Jahanshahi}},
  \bibinfo {author} {\bibfnamefont {A.}~\bibnamefont {Ldov}}, \ and\ \bibinfo
  {author} {\bibfnamefont {H.}~\bibnamefont {Löwen}},\ }\href {\doibase
  10.1038/s41467-018-07596-x} {\bibfield  {journal} {\bibinfo  {journal}
  {Nature Communications}\ }\textbf {\bibinfo {volume} {9}},\ \bibinfo {pages}
  {5156} (\bibinfo {year} {2018})}\BibitemShut {NoStop}%
\bibitem [{\citenamefont {Rogers}\ \emph {et~al.}(2021)\citenamefont {Rogers},
  \citenamefont {Kim}, \citenamefont {Li}, \citenamefont {Kim}, \citenamefont
  {Park}, \citenamefont {Jang}, \citenamefont {Wang}, \citenamefont {Xie},
  \citenamefont {Won}, \citenamefont {Jang}, \citenamefont {Lee}, \citenamefont
  {Chung}, \citenamefont {Jung}, \citenamefont {Heo}, \citenamefont {Lee},
  \citenamefont {Kim}, \citenamefont {Tengfei}, \citenamefont {Kim},
  \citenamefont {Prasopsukh},\ and\ \citenamefont {Zhang}}]{3Dmicroflyer}%
  \BibitemOpen
  \bibfield  {author} {\bibinfo {author} {\bibfnamefont {J.}~\bibnamefont
  {Rogers}}, \bibinfo {author} {\bibfnamefont {B.~H.}\ \bibnamefont {Kim}},
  \bibinfo {author} {\bibfnamefont {K.}~\bibnamefont {Li}}, \bibinfo {author}
  {\bibfnamefont {J.-T.}\ \bibnamefont {Kim}}, \bibinfo {author} {\bibfnamefont
  {Y.}~\bibnamefont {Park}}, \bibinfo {author} {\bibfnamefont {H.}~\bibnamefont
  {Jang}}, \bibinfo {author} {\bibfnamefont {X.}~\bibnamefont {Wang}}, \bibinfo
  {author} {\bibfnamefont {Z.}~\bibnamefont {Xie}}, \bibinfo {author}
  {\bibfnamefont {S.}~\bibnamefont {Won}}, \bibinfo {author} {\bibfnamefont
  {W.}~\bibnamefont {Jang}}, \bibinfo {author} {\bibfnamefont {K.}~\bibnamefont
  {Lee}}, \bibinfo {author} {\bibfnamefont {T.}~\bibnamefont {Chung}}, \bibinfo
  {author} {\bibfnamefont {Y.~H.}\ \bibnamefont {Jung}}, \bibinfo {author}
  {\bibfnamefont {S.}~\bibnamefont {Heo}}, \bibinfo {author} {\bibfnamefont
  {Y.}~\bibnamefont {Lee}}, \bibinfo {author} {\bibfnamefont {J.}~\bibnamefont
  {Kim}}, \bibinfo {author} {\bibfnamefont {C.}~\bibnamefont {Tengfei}},
  \bibinfo {author} {\bibfnamefont {Y.}~\bibnamefont {Kim}}, \bibinfo {author}
  {\bibfnamefont {P.}~\bibnamefont {Prasopsukh}}, \ and\ \bibinfo {author}
  {\bibfnamefont {Y.}~\bibnamefont {Zhang}},\ }\href {\doibase
  10.21203/rs.3.rs-142068/v1} {\  (\bibinfo {year} {2021}),\
  10.21203/rs.3.rs-142068/v1}\BibitemShut {NoStop}%
\bibitem [{\citenamefont {Lowen}(2020)}]{lowen2020}%
  \BibitemOpen
  \bibfield  {author} {\bibinfo {author} {\bibfnamefont {H.}~\bibnamefont
  {Lowen}},\ }\href {\doibase 10.1063/1.5134455} {\bibfield  {journal}
  {\bibinfo  {journal} {The Journal of Chemical Physics}\ }\textbf {\bibinfo
  {volume} {152}},\ \bibinfo {pages} {040901} (\bibinfo {year}
  {2020})}\BibitemShut {NoStop}%
\bibitem [{\citenamefont {Safford}\ \emph {et~al.}(2009)\citenamefont
  {Safford}, \citenamefont {Kantor}, \citenamefont {Kardar},\ and\
  \citenamefont {Kudrolli}}]{granular_chain}%
  \BibitemOpen
  \bibfield  {author} {\bibinfo {author} {\bibfnamefont {K.}~\bibnamefont
  {Safford}}, \bibinfo {author} {\bibfnamefont {Y.}~\bibnamefont {Kantor}},
  \bibinfo {author} {\bibfnamefont {M.}~\bibnamefont {Kardar}}, \ and\ \bibinfo
  {author} {\bibfnamefont {A.}~\bibnamefont {Kudrolli}},\ }\href {\doibase
  10.1103/PhysRevE.79.061304} {\bibfield  {journal} {\bibinfo  {journal} {Phys.
  Rev. E}\ }\textbf {\bibinfo {volume} {79}},\ \bibinfo {pages} {061304}
  (\bibinfo {year} {2009})}\BibitemShut {NoStop}%
\bibitem [{\citenamefont {Malakar}\ \emph {et~al.}(2020)\citenamefont
  {Malakar}, \citenamefont {Das}, \citenamefont {Kundu}, \citenamefont
  {Kumar},\ and\ \citenamefont {Dhar}}]{Malakar}%
  \BibitemOpen
  \bibfield  {author} {\bibinfo {author} {\bibfnamefont {K.}~\bibnamefont
  {Malakar}}, \bibinfo {author} {\bibfnamefont {A.}~\bibnamefont {Das}},
  \bibinfo {author} {\bibfnamefont {A.}~\bibnamefont {Kundu}}, \bibinfo
  {author} {\bibfnamefont {K.~V.}\ \bibnamefont {Kumar}}, \ and\ \bibinfo
  {author} {\bibfnamefont {A.}~\bibnamefont {Dhar}},\ }\href {\doibase
  10.1103/PhysRevE.101.022610} {\bibfield  {journal} {\bibinfo  {journal}
  {Phys. Rev. E}\ }\textbf {\bibinfo {volume} {101}},\ \bibinfo {pages}
  {022610} (\bibinfo {year} {2020})}\BibitemShut {NoStop}%
\bibitem [{\citenamefont {Jahanshahi}(2019)}]{jahanshahi}%
  \BibitemOpen
  \bibfield  {author} {\bibinfo {author} {\bibfnamefont {S.}~\bibnamefont
  {Jahanshahi}},\ }\emph {\bibinfo {title} {Microswimmers and microflyers in
  various complex environments}},\ \href
  {https://docserv.uni-duesseldorf.de/servlets/DocumentServlet?id=51830} {Ph.D.
  thesis} (\bibinfo {year} {2019})\BibitemShut {NoStop}%
\bibitem [{\citenamefont {Chatterjee}\ \emph {et~al.}(2021)\citenamefont
  {Chatterjee}, \citenamefont {Rana}, \citenamefont {Simha}, \citenamefont
  {Perlekar},\ and\ \citenamefont {Ramaswamy}}]{inertial_flock_21}%
  \BibitemOpen
  \bibfield  {author} {\bibinfo {author} {\bibfnamefont {R.}~\bibnamefont
  {Chatterjee}}, \bibinfo {author} {\bibfnamefont {N.}~\bibnamefont {Rana}},
  \bibinfo {author} {\bibfnamefont {R.~A.}\ \bibnamefont {Simha}}, \bibinfo
  {author} {\bibfnamefont {P.}~\bibnamefont {Perlekar}}, \ and\ \bibinfo
  {author} {\bibfnamefont {S.}~\bibnamefont {Ramaswamy}},\ }\href {\doibase
  10.1103/PhysRevX.11.031063} {\bibfield  {journal} {\bibinfo  {journal} {Phys.
  Rev. X}\ }\textbf {\bibinfo {volume} {11}},\ \bibinfo {pages} {031063}
  (\bibinfo {year} {2021})}\BibitemShut {NoStop}%
\bibitem [{\citenamefont {Caprini}\ and\ \citenamefont {Marini
  Bettolo~Marconi}(2021)}]{Inertial_AOUP}%
  \BibitemOpen
  \bibfield  {author} {\bibinfo {author} {\bibfnamefont {L.}~\bibnamefont
  {Caprini}}\ and\ \bibinfo {author} {\bibfnamefont {U.}~\bibnamefont {Marini
  Bettolo~Marconi}},\ }\href {\doibase 10.1063/5.0030940} {\bibfield  {journal}
  {\bibinfo  {journal} {The Journal of Chemical Physics}\ }\textbf {\bibinfo
  {volume} {154}},\ \bibinfo {pages} {024902} (\bibinfo {year}
  {2021})}\BibitemShut {NoStop}%
\bibitem [{\citenamefont {Sandoval}(2020)}]{ABP_diffusion}%
  \BibitemOpen
  \bibfield  {author} {\bibinfo {author} {\bibfnamefont {M.}~\bibnamefont
  {Sandoval}},\ }\href {\doibase 10.1103/PhysRevE.101.012606} {\bibfield
  {journal} {\bibinfo  {journal} {Phys. Rev. E}\ }\textbf {\bibinfo {volume}
  {101}},\ \bibinfo {pages} {012606} (\bibinfo {year} {2020})}\BibitemShut
  {NoStop}%
\bibitem [{\citenamefont {Gutierrez-Martinez}\ and\ \citenamefont
  {Sandoval}(2020)}]{trappped_ABP_inertia}%
  \BibitemOpen
  \bibfield  {author} {\bibinfo {author} {\bibfnamefont {L.~L.}\ \bibnamefont
  {Gutierrez-Martinez}}\ and\ \bibinfo {author} {\bibfnamefont
  {M.}~\bibnamefont {Sandoval}},\ }\href {\doibase 10.1063/5.0011270}
  {\bibfield  {journal} {\bibinfo  {journal} {The Journal of Chemical Physics}\
  }\textbf {\bibinfo {volume} {153}},\ \bibinfo {pages} {044906} (\bibinfo
  {year} {2020})}\BibitemShut {NoStop}%
\bibitem [{\citenamefont {Dai}\ \emph {et~al.}(2020)\citenamefont {Dai},
  \citenamefont {Bruss},\ and\ \citenamefont {Glotzer}}]{Dai_inertia}%
  \BibitemOpen
  \bibfield  {author} {\bibinfo {author} {\bibfnamefont {C.}~\bibnamefont
  {Dai}}, \bibinfo {author} {\bibfnamefont {I.~R.}\ \bibnamefont {Bruss}}, \
  and\ \bibinfo {author} {\bibfnamefont {S.~C.}\ \bibnamefont {Glotzer}},\
  }\href {\doibase 10.1039/C9SM01683J} {\bibfield  {journal} {\bibinfo
  {journal} {Soft Matter}\ }\textbf {\bibinfo {volume} {16}},\ \bibinfo {pages}
  {2847} (\bibinfo {year} {2020})}\BibitemShut {NoStop}%
\bibitem [{\citenamefont {Sprenger}\ \emph {et~al.}(2021)\citenamefont
  {Sprenger}, \citenamefont {Jahanshahi}, \citenamefont {Ivlev},\ and\
  \citenamefont {L\"owen}}]{Ineria_time_dependent}%
  \BibitemOpen
  \bibfield  {author} {\bibinfo {author} {\bibfnamefont {A.~R.}\ \bibnamefont
  {Sprenger}}, \bibinfo {author} {\bibfnamefont {S.}~\bibnamefont
  {Jahanshahi}}, \bibinfo {author} {\bibfnamefont {A.~V.}\ \bibnamefont
  {Ivlev}}, \ and\ \bibinfo {author} {\bibfnamefont {H.}~\bibnamefont
  {L\"owen}},\ }\href {\doibase 10.1103/PhysRevE.103.042601} {\bibfield
  {journal} {\bibinfo  {journal} {Phys. Rev. E}\ }\textbf {\bibinfo {volume}
  {103}},\ \bibinfo {pages} {042601} (\bibinfo {year} {2021})}\BibitemShut
  {NoStop}%
\bibitem [{\citenamefont {Nguyen}\ \emph
  {et~al.}(2021{\natexlab{a}})\citenamefont {Nguyen}, \citenamefont
  {Wittmann},\ and\ \citenamefont {Löwen}}]{lowen-AOUP}%
  \BibitemOpen
  \bibfield  {author} {\bibinfo {author} {\bibfnamefont {G.~H.}\ \bibnamefont
  {Nguyen}}, \bibinfo {author} {\bibfnamefont {R.}~\bibnamefont {Wittmann}}, \
  and\ \bibinfo {author} {\bibfnamefont {H.}~\bibnamefont {Löwen}},\ }\href
  {\doibase 10.1088/1361-648x/ac2c3f} {\bibfield  {journal} {\bibinfo
  {journal} {Journal of Physics: Condensed Matter}\ }\textbf {\bibinfo {volume}
  {34}},\ \bibinfo {pages} {035101} (\bibinfo {year}
  {2021}{\natexlab{a}})}\BibitemShut {NoStop}%
\bibitem [{\citenamefont {Caprini}\ \emph {et~al.}(2022)\citenamefont
  {Caprini}, \citenamefont {Gupta},\ and\ \citenamefont
  {Löwen}}]{caprini2022role}%
  \BibitemOpen
  \bibfield  {author} {\bibinfo {author} {\bibfnamefont {L.}~\bibnamefont
  {Caprini}}, \bibinfo {author} {\bibfnamefont {R.~K.}\ \bibnamefont {Gupta}},
  \ and\ \bibinfo {author} {\bibfnamefont {H.}~\bibnamefont {Löwen}},\
  }\href@noop {} {\enquote {\bibinfo {title} {Role of rotational inertia for
  collective phenomena in active matter},}\ } (\bibinfo {year} {2022}),\
  \Eprint {http://arxiv.org/abs/2206.14324} {arXiv:2206.14324 [cond-mat.soft]}
  \BibitemShut {NoStop}%
\bibitem [{\citenamefont {Winkler}\ \emph {et~al.}(2017)\citenamefont
  {Winkler}, \citenamefont {Elgeti},\ and\ \citenamefont
  {Gompper}}]{active_poly_review}%
  \BibitemOpen
  \bibfield  {author} {\bibinfo {author} {\bibfnamefont {R.~G.}\ \bibnamefont
  {Winkler}}, \bibinfo {author} {\bibfnamefont {J.}~\bibnamefont {Elgeti}}, \
  and\ \bibinfo {author} {\bibfnamefont {G.}~\bibnamefont {Gompper}},\ }\href
  {\doibase 10.7566/JPSJ.86.101014} {\bibfield  {journal} {\bibinfo  {journal}
  {Journal of the Physical Society of Japan}\ }\textbf {\bibinfo {volume}
  {86}},\ \bibinfo {pages} {101014} (\bibinfo {year} {2017})}\BibitemShut
  {NoStop}%
\bibitem [{\citenamefont {Winkler}\ and\ \citenamefont
  {Gompper}(2020)}]{active_pol_20}%
  \BibitemOpen
  \bibfield  {author} {\bibinfo {author} {\bibfnamefont {R.~G.}\ \bibnamefont
  {Winkler}}\ and\ \bibinfo {author} {\bibfnamefont {G.}~\bibnamefont
  {Gompper}},\ }\href {\doibase 10.1063/5.0011466} {\bibfield  {journal}
  {\bibinfo  {journal} {The Journal of Chemical Physics}\ }\textbf {\bibinfo
  {volume} {153}},\ \bibinfo {pages} {040901} (\bibinfo {year}
  {2020})}\BibitemShut {NoStop}%
\bibitem [{\citenamefont {Alberts}(2007)}]{biopolymers1}%
  \BibitemOpen
  \bibfield  {author} {\bibinfo {author} {\bibfnamefont {B.}~\bibnamefont
  {Alberts}},\ }\href@noop {} {\emph {\bibinfo {title} {Molecular biology of
  the cell}}}\ (\bibinfo  {publisher} {TPB},\ \bibinfo {year}
  {2007})\BibitemShut {NoStop}%
\bibitem [{\citenamefont {Bray}(2001)}]{biopolymers2}%
  \BibitemOpen
  \bibfield  {author} {\bibinfo {author} {\bibfnamefont {D.}~\bibnamefont
  {Bray}},\ }\href@noop {} {\emph {\bibinfo {title} {Cell Movements: From
  Molecules to Motility}}}\ (\bibinfo  {publisher} {Garland Science, New York,
  2nd ed.},\ \bibinfo {year} {2001})\BibitemShut {NoStop}%
\bibitem [{\citenamefont {Deblais}\ \emph
  {et~al.}(2020{\natexlab{a}})\citenamefont {Deblais}, \citenamefont {Maggs},
  \citenamefont {Bonn},\ and\ \citenamefont {Woutersen}}]{aquarium_worms}%
  \BibitemOpen
  \bibfield  {author} {\bibinfo {author} {\bibfnamefont {A.}~\bibnamefont
  {Deblais}}, \bibinfo {author} {\bibfnamefont {A.~C.}\ \bibnamefont {Maggs}},
  \bibinfo {author} {\bibfnamefont {D.}~\bibnamefont {Bonn}}, \ and\ \bibinfo
  {author} {\bibfnamefont {S.}~\bibnamefont {Woutersen}},\ }\href {\doibase
  10.1103/PhysRevLett.124.208006} {\bibfield  {journal} {\bibinfo  {journal}
  {Phys. Rev. Lett.}\ }\textbf {\bibinfo {volume} {124}},\ \bibinfo {pages}
  {208006} (\bibinfo {year} {2020}{\natexlab{a}})}\BibitemShut {NoStop}%
\bibitem [{\citenamefont {Deblais}\ \emph
  {et~al.}(2020{\natexlab{b}})\citenamefont {Deblais}, \citenamefont
  {Woutersen},\ and\ \citenamefont {Bonn}}]{worms_rheology}%
  \BibitemOpen
  \bibfield  {author} {\bibinfo {author} {\bibfnamefont {A.}~\bibnamefont
  {Deblais}}, \bibinfo {author} {\bibfnamefont {S.}~\bibnamefont {Woutersen}},
  \ and\ \bibinfo {author} {\bibfnamefont {D.}~\bibnamefont {Bonn}},\ }\href
  {\doibase 10.1103/PhysRevLett.124.188002} {\bibfield  {journal} {\bibinfo
  {journal} {Phys. Rev. Lett.}\ }\textbf {\bibinfo {volume} {124}},\ \bibinfo
  {pages} {188002} (\bibinfo {year} {2020}{\natexlab{b}})}\BibitemShut
  {NoStop}%
\bibitem [{\citenamefont {Ozkan~Aydin}\ \emph {et~al.}(2021)\citenamefont
  {Ozkan~Aydin}, \citenamefont {Goldman},\ and\ \citenamefont
  {Bhamla}}]{California_worms}%
  \BibitemOpen
  \bibfield  {author} {\bibinfo {author} {\bibfnamefont {Y.}~\bibnamefont
  {Ozkan~Aydin}}, \bibinfo {author} {\bibfnamefont {D.}~\bibnamefont
  {Goldman}}, \ and\ \bibinfo {author} {\bibfnamefont {S.}~\bibnamefont
  {Bhamla}},\ }\href {\doibase 10.1073/pnas.2010542118} {\bibfield  {journal}
  {\bibinfo  {journal} {Proceedings of the National Academy of Sciences}\
  }\textbf {\bibinfo {volume} {118}},\ \bibinfo {pages} {e2010542118} (\bibinfo
  {year} {2021})}\BibitemShut {NoStop}%
\bibitem [{\citenamefont {Marvi}\ \emph {et~al.}(2014)\citenamefont {Marvi},
  \citenamefont {Gong}, \citenamefont {Gravish}, \citenamefont {Astley},
  \citenamefont {Travers}, \citenamefont {Hatton}, \citenamefont {Mendelson},
  \citenamefont {Choset}, \citenamefont {Hu},\ and\ \citenamefont
  {Goldman}}]{snakes}%
  \BibitemOpen
  \bibfield  {author} {\bibinfo {author} {\bibfnamefont {H.}~\bibnamefont
  {Marvi}}, \bibinfo {author} {\bibfnamefont {C.}~\bibnamefont {Gong}},
  \bibinfo {author} {\bibfnamefont {N.}~\bibnamefont {Gravish}}, \bibinfo
  {author} {\bibfnamefont {H.}~\bibnamefont {Astley}}, \bibinfo {author}
  {\bibfnamefont {M.}~\bibnamefont {Travers}}, \bibinfo {author} {\bibfnamefont
  {R.~L.}\ \bibnamefont {Hatton}}, \bibinfo {author} {\bibfnamefont {J.~R.}\
  \bibnamefont {Mendelson}}, \bibinfo {author} {\bibfnamefont {H.}~\bibnamefont
  {Choset}}, \bibinfo {author} {\bibfnamefont {D.~L.}\ \bibnamefont {Hu}}, \
  and\ \bibinfo {author} {\bibfnamefont {D.~I.}\ \bibnamefont {Goldman}},\
  }\href {\doibase 10.1126/science.1255718} {\bibfield  {journal} {\bibinfo
  {journal} {Science}\ }\textbf {\bibinfo {volume} {346}},\ \bibinfo {pages}
  {224} (\bibinfo {year} {2014})}\BibitemShut {NoStop}%
\bibitem [{\citenamefont {Zheng}\ \emph {et~al.}(2021)\citenamefont {Zheng},
  \citenamefont {Brandenbourger}, \citenamefont {Robinet}, \citenamefont
  {Schall}, \citenamefont {Lerner},\ and\ \citenamefont {Coulais}}]{Corentin}%
  \BibitemOpen
  \bibfield  {author} {\bibinfo {author} {\bibfnamefont {E.}~\bibnamefont
  {Zheng}}, \bibinfo {author} {\bibfnamefont {M.}~\bibnamefont
  {Brandenbourger}}, \bibinfo {author} {\bibfnamefont {L.}~\bibnamefont
  {Robinet}}, \bibinfo {author} {\bibfnamefont {P.}~\bibnamefont {Schall}},
  \bibinfo {author} {\bibfnamefont {E.}~\bibnamefont {Lerner}}, \ and\ \bibinfo
  {author} {\bibfnamefont {C.}~\bibnamefont {Coulais}},\ }\href {\doibase
  10.48550/ARXIV.2106.05721} {\enquote {\bibinfo {title} {Self-oscillation and
  synchronisation transitions in elasto-active structures},}\ } (\bibinfo
  {year} {2021})\BibitemShut {NoStop}%
\bibitem [{\citenamefont {Isele-Holder}\ \emph {et~al.}(2015)\citenamefont
  {Isele-Holder}, \citenamefont {Elgeti},\ and\ \citenamefont
  {Gompper}}]{Isele}%
  \BibitemOpen
  \bibfield  {author} {\bibinfo {author} {\bibfnamefont {R.~E.}\ \bibnamefont
  {Isele-Holder}}, \bibinfo {author} {\bibfnamefont {J.}~\bibnamefont
  {Elgeti}}, \ and\ \bibinfo {author} {\bibfnamefont {G.}~\bibnamefont
  {Gompper}},\ }\href {\doibase 10.1039/C5SM01683E} {\bibfield  {journal}
  {\bibinfo  {journal} {Soft Matter}\ }\textbf {\bibinfo {volume} {11}},\
  \bibinfo {pages} {7181} (\bibinfo {year} {2015})}\BibitemShut {NoStop}%
\bibitem [{\citenamefont {Bianco}\ \emph {et~al.}(2018)\citenamefont {Bianco},
  \citenamefont {Locatelli},\ and\ \citenamefont
  {Malgaretti}}]{Active_flexible}%
  \BibitemOpen
  \bibfield  {author} {\bibinfo {author} {\bibfnamefont {V.}~\bibnamefont
  {Bianco}}, \bibinfo {author} {\bibfnamefont {E.}~\bibnamefont {Locatelli}}, \
  and\ \bibinfo {author} {\bibfnamefont {P.}~\bibnamefont {Malgaretti}},\
  }\href {\doibase 10.1103/PhysRevLett.121.217802} {\bibfield  {journal}
  {\bibinfo  {journal} {Phys. Rev. Lett.}\ }\textbf {\bibinfo {volume} {121}},\
  \bibinfo {pages} {217802} (\bibinfo {year} {2018})}\BibitemShut {NoStop}%
\bibitem [{\citenamefont {Anand}\ and\ \citenamefont
  {Singh}(2018)}]{active_priodic_force}%
  \BibitemOpen
  \bibfield  {author} {\bibinfo {author} {\bibfnamefont {S.~K.}\ \bibnamefont
  {Anand}}\ and\ \bibinfo {author} {\bibfnamefont {S.~P.}\ \bibnamefont
  {Singh}},\ }\href {\doibase 10.1103/PhysRevE.98.042501} {\bibfield  {journal}
  {\bibinfo  {journal} {Phys. Rev. E}\ }\textbf {\bibinfo {volume} {98}},\
  \bibinfo {pages} {042501} (\bibinfo {year} {2018})}\BibitemShut {NoStop}%
\bibitem [{\citenamefont {Mokhtari}\ and\ \citenamefont
  {Zippelius}(2019)}]{Mokhtari}%
  \BibitemOpen
  \bibfield  {author} {\bibinfo {author} {\bibfnamefont {Z.}~\bibnamefont
  {Mokhtari}}\ and\ \bibinfo {author} {\bibfnamefont {A.}~\bibnamefont
  {Zippelius}},\ }\href {\doibase 10.1103/PhysRevLett.123.028001} {\bibfield
  {journal} {\bibinfo  {journal} {Phys. Rev. Lett.}\ }\textbf {\bibinfo
  {volume} {123}},\ \bibinfo {pages} {028001} (\bibinfo {year}
  {2019})}\BibitemShut {NoStop}%
\bibitem [{\citenamefont {Ghosh}\ and\ \citenamefont
  {Spakowitz}(2022)}]{Ghosh_22}%
  \BibitemOpen
  \bibfield  {author} {\bibinfo {author} {\bibfnamefont {A.}~\bibnamefont
  {Ghosh}}\ and\ \bibinfo {author} {\bibfnamefont {A.~J.}\ \bibnamefont
  {Spakowitz}},\ }\href {\doibase 10.1039/D2SM00593J} {\bibfield  {journal}
  {\bibinfo  {journal} {Soft Matter}\ }\textbf {\bibinfo {volume} {18}},\
  \bibinfo {pages} {6629} (\bibinfo {year} {2022})}\BibitemShut {NoStop}%
\bibitem [{\citenamefont {Eisenstecken}\ \emph {et~al.}(2016)\citenamefont
  {Eisenstecken}, \citenamefont {Gompper},\ and\ \citenamefont
  {Winkler}}]{ABPO_ANALYTICS-eisenstecken_gompper_winkler_2016}%
  \BibitemOpen
  \bibfield  {author} {\bibinfo {author} {\bibfnamefont {T.}~\bibnamefont
  {Eisenstecken}}, \bibinfo {author} {\bibfnamefont {G.}~\bibnamefont
  {Gompper}}, \ and\ \bibinfo {author} {\bibfnamefont {R.}~\bibnamefont
  {Winkler}},\ }\href {\doibase 10.3390/polym8080304} {\bibfield  {journal}
  {\bibinfo  {journal} {Polymers}\ }\textbf {\bibinfo {volume} {8}},\ \bibinfo
  {pages} {304} (\bibinfo {year} {2016})}\BibitemShut {NoStop}%
\bibitem [{\citenamefont {Philipps}\ \emph {et~al.}(2022)\citenamefont
  {Philipps}, \citenamefont {Gompper},\ and\ \citenamefont
  {Winkler}}]{Phillipps_22}%
  \BibitemOpen
  \bibfield  {author} {\bibinfo {author} {\bibfnamefont {C.~A.}\ \bibnamefont
  {Philipps}}, \bibinfo {author} {\bibfnamefont {G.}~\bibnamefont {Gompper}}, \
  and\ \bibinfo {author} {\bibfnamefont {R.~G.}\ \bibnamefont {Winkler}},\
  }\href {https://doi.org/10.1063/5.0120493} {\bibfield  {journal} {\bibinfo
  {journal} {The Journal of Chemical Physics}\ }\textbf {\bibinfo {volume} {0}}
  (\bibinfo {year} {2022})}\BibitemShut {NoStop}%
\bibitem [{\citenamefont {Nguyen}\ \emph
  {et~al.}(2021{\natexlab{b}})\citenamefont {Nguyen}, \citenamefont
  {Ozkan-Aydin}, \citenamefont {Tuazon}, \citenamefont {Goldman}, \citenamefont
  {Bhamla},\ and\ \citenamefont {Peleg}}]{nguyen2021emergent}%
  \BibitemOpen
  \bibfield  {author} {\bibinfo {author} {\bibfnamefont {C.}~\bibnamefont
  {Nguyen}}, \bibinfo {author} {\bibfnamefont {Y.}~\bibnamefont {Ozkan-Aydin}},
  \bibinfo {author} {\bibfnamefont {H.}~\bibnamefont {Tuazon}}, \bibinfo
  {author} {\bibfnamefont {D.~I.}\ \bibnamefont {Goldman}}, \bibinfo {author}
  {\bibfnamefont {M.~S.}\ \bibnamefont {Bhamla}}, \ and\ \bibinfo {author}
  {\bibfnamefont {O.}~\bibnamefont {Peleg}},\ }\href@noop {} {\bibfield
  {journal} {\bibinfo  {journal} {Frontiers in Physics}\ }\textbf {\bibinfo
  {volume} {9}} (\bibinfo {year} {2021}{\natexlab{b}})}\BibitemShut {NoStop}%
\bibitem [{\citenamefont {Luo}\ \emph {et~al.}(2014)\citenamefont {Luo},
  \citenamefont {Liu}, \citenamefont {Gao},\ and\ \citenamefont
  {Lu}}]{snakerobot2}%
  \BibitemOpen
  \bibfield  {author} {\bibinfo {author} {\bibfnamefont {Y.}~\bibnamefont
  {Luo}}, \bibinfo {author} {\bibfnamefont {J.}~\bibnamefont {Liu}}, \bibinfo
  {author} {\bibfnamefont {Y.}~\bibnamefont {Gao}}, \ and\ \bibinfo {author}
  {\bibfnamefont {Z.}~\bibnamefont {Lu}},\ }in\ \href@noop {} {\emph {\bibinfo
  {booktitle} {Intelligent Robotics and Applications}}},\ \bibinfo {editor}
  {edited by\ \bibinfo {editor} {\bibfnamefont {X.}~\bibnamefont {Zhang}},
  \bibinfo {editor} {\bibfnamefont {H.}~\bibnamefont {Liu}}, \bibinfo {editor}
  {\bibfnamefont {Z.}~\bibnamefont {Chen}}, \ and\ \bibinfo {editor}
  {\bibfnamefont {N.}~\bibnamefont {Wang}}}\ (\bibinfo  {publisher} {Springer
  International Publishing},\ \bibinfo {address} {Cham},\ \bibinfo {year}
  {2014})\ pp.\ \bibinfo {pages} {352--363}\BibitemShut {NoStop}%
\bibitem [{\citenamefont {Joesinstructables}\ and\ \citenamefont
  {Instructables}(2017)}]{snakerobot1}%
  \BibitemOpen
  \bibfield  {author} {\bibinfo {author} {\bibnamefont {Joesinstructables}}\
  and\ \bibinfo {author} {\bibnamefont {Instructables}},\ }\href
  {https://www.instructables.com/Snake-Robot-1/} {\enquote {\bibinfo {title}
  {Snake robot},}\ } (\bibinfo {year} {2017})\BibitemShut {NoStop}%
\bibitem [{\citenamefont {Wong}\ and\ \citenamefont
  {Choi}(2018)}]{inertia_Rouse}%
  \BibitemOpen
  \bibfield  {author} {\bibinfo {author} {\bibfnamefont {C.~P.~J.}\
  \bibnamefont {Wong}}\ and\ \bibinfo {author} {\bibfnamefont {P.}~\bibnamefont
  {Choi}},\ }\href {\doibase https://doi.org/10.1016/j.commatsci.2018.08.042}
  {\bibfield  {journal} {\bibinfo  {journal} {Computational Materials Science}\
  }\textbf {\bibinfo {volume} {155}},\ \bibinfo {pages} {320} (\bibinfo {year}
  {2018})}\BibitemShut {NoStop}%
\bibitem [{\citenamefont {Deutsch}(2012)}]{Deutsch-review}%
  \BibitemOpen
  \bibfield  {author} {\bibinfo {author} {\bibfnamefont {J.~M.}\ \bibnamefont
  {Deutsch}},\ }\href {\doibase https://doi.org/10.1002/polb.23040} {\bibfield
  {journal} {\bibinfo  {journal} {Journal of Polymer Science Part B: Polymer
  Physics}\ }\textbf {\bibinfo {volume} {50}},\ \bibinfo {pages} {379}
  (\bibinfo {year} {2012})}\BibitemShut {NoStop}%
\bibitem [{\citenamefont {Deutsch}(2011)}]{Deutsch}%
  \BibitemOpen
  \bibfield  {author} {\bibinfo {author} {\bibfnamefont {J.~M.}\ \bibnamefont
  {Deutsch}},\ }\href {\doibase 10.1103/PhysRevE.83.051801} {\bibfield
  {journal} {\bibinfo  {journal} {Phys. Rev. E}\ }\textbf {\bibinfo {volume}
  {83}},\ \bibinfo {pages} {051801} (\bibinfo {year} {2011})}\BibitemShut
  {NoStop}%
\bibitem [{\citenamefont {Weeks}\ \emph {et~al.}(1971)\citenamefont {Weeks},
  \citenamefont {Chandler},\ and\ \citenamefont {Andersen}}]{WCA}%
  \BibitemOpen
  \bibfield  {author} {\bibinfo {author} {\bibfnamefont {J.~D.}\ \bibnamefont
  {Weeks}}, \bibinfo {author} {\bibfnamefont {D.}~\bibnamefont {Chandler}}, \
  and\ \bibinfo {author} {\bibfnamefont {H.~C.}\ \bibnamefont {Andersen}},\
  }\href {\doibase 10.1063/1.1674820} {\bibfield  {journal} {\bibinfo
  {journal} {The Journal of Chemical Physics}\ }\textbf {\bibinfo {volume}
  {54}},\ \bibinfo {pages} {5237} (\bibinfo {year} {1971})}\BibitemShut
  {NoStop}%
\bibitem [{SI()}]{SI}%
  \BibitemOpen
  \href@noop {} {\enquote {\bibinfo {title} {See supplemental material at url
  for a sample video, additional numerical data, and detailed theoretical
  calculations.}}\ }\BibitemShut {NoStop}%
\bibitem [{\citenamefont {Rudnick}\ and\ \citenamefont
  {Gaspari}(1986)}]{asphericity}%
  \BibitemOpen
  \bibfield  {author} {\bibinfo {author} {\bibfnamefont {J.}~\bibnamefont
  {Rudnick}}\ and\ \bibinfo {author} {\bibfnamefont {G.}~\bibnamefont
  {Gaspari}},\ }\href {\doibase 10.1088/0305-4470/19/4/004} {\bibfield
  {journal} {\bibinfo  {journal} {Journal of Physics A: Mathematical and
  General}\ }\textbf {\bibinfo {volume} {19}},\ \bibinfo {pages} {L191}
  (\bibinfo {year} {1986})}\BibitemShut {NoStop}%
\bibitem [{\citenamefont {Barkema}\ \emph {et~al.}(2014)\citenamefont
  {Barkema}, \citenamefont {Panja},\ and\ \citenamefont {van
  Leeuwen}}]{Barkema2014}%
  \BibitemOpen
  \bibfield  {author} {\bibinfo {author} {\bibfnamefont {G.~T.}\ \bibnamefont
  {Barkema}}, \bibinfo {author} {\bibfnamefont {D.}~\bibnamefont {Panja}}, \
  and\ \bibinfo {author} {\bibfnamefont {J.~M.~J.}\ \bibnamefont {van
  Leeuwen}},\ }\href {http://stacks.iop.org/1742-5468/2014/i=11/a=P11008}
  {\bibfield  {journal} {\bibinfo  {journal} {Journal of Statistical Mechanics:
  Theory and Experiment}\ }\textbf {\bibinfo {volume} {2014}},\ \bibinfo
  {pages} {P11008} (\bibinfo {year} {2014})}\BibitemShut {NoStop}%
\bibitem [{\citenamefont {Doi}\ and\ \citenamefont
  {Edwards}(1986)}]{polymerDoi}%
  \BibitemOpen
  \bibfield  {author} {\bibinfo {author} {\bibfnamefont {M.}~\bibnamefont
  {Doi}}\ and\ \bibinfo {author} {\bibfnamefont {S.~F.}\ \bibnamefont
  {Edwards}},\ }\href@noop {} {\emph {\bibinfo {title} {The Theory of Polymer
  Dynamics}}}\ (\bibinfo  {publisher} {Clarendon Press, Oxford},\ \bibinfo
  {year} {1986})\BibitemShut {NoStop}%
\bibitem [{\citenamefont {Martín-Gómez}\ \emph {et~al.}(2019)\citenamefont
  {Martín-Gómez}, \citenamefont {Eisenstecken}, \citenamefont {Gompper},\
  and\ \citenamefont {Winkler}}]{ABPO-N200}%
  \BibitemOpen
  \bibfield  {author} {\bibinfo {author} {\bibfnamefont {A.}~\bibnamefont
  {Martín-Gómez}}, \bibinfo {author} {\bibfnamefont {T.}~\bibnamefont
  {Eisenstecken}}, \bibinfo {author} {\bibfnamefont {G.}~\bibnamefont
  {Gompper}}, \ and\ \bibinfo {author} {\bibfnamefont {R.~G.}\ \bibnamefont
  {Winkler}},\ }\href {\doibase 10.1039/c9sm00391f} {\bibfield  {journal}
  {\bibinfo  {journal} {Soft Matter}\ }\textbf {\bibinfo {volume} {15}},\
  \bibinfo {pages} {3957–3969} (\bibinfo {year} {2019})}\BibitemShut
  {NoStop}%
\bibitem [{\citenamefont {Peterson}\ \emph {et~al.}(2020)\citenamefont
  {Peterson}, \citenamefont {Hagan},\ and\ \citenamefont
  {Baskaran}}]{BASKARAN}%
  \BibitemOpen
  \bibfield  {author} {\bibinfo {author} {\bibfnamefont {M.~S.~E.}\
  \bibnamefont {Peterson}}, \bibinfo {author} {\bibfnamefont {M.~F.}\
  \bibnamefont {Hagan}}, \ and\ \bibinfo {author} {\bibfnamefont
  {A.}~\bibnamefont {Baskaran}},\ }\href {\doibase 10.1088/1742-5468/ab6097}
  {\bibfield  {journal} {\bibinfo  {journal} {Journal of Statistical Mechanics:
  Theory and Experiment}\ }\textbf {\bibinfo {volume} {2020}},\ \bibinfo
  {pages} {013216} (\bibinfo {year} {2020})}\BibitemShut {NoStop}%
\bibitem [{\citenamefont {Locatelli}\ \emph {et~al.}(2021)\citenamefont
  {Locatelli}, \citenamefont {Bianco},\ and\ \citenamefont
  {Malgaretti}}]{Locatelli_21}%
  \BibitemOpen
  \bibfield  {author} {\bibinfo {author} {\bibfnamefont {E.}~\bibnamefont
  {Locatelli}}, \bibinfo {author} {\bibfnamefont {V.}~\bibnamefont {Bianco}}, \
  and\ \bibinfo {author} {\bibfnamefont {P.}~\bibnamefont {Malgaretti}},\
  }\href {\doibase 10.1103/PhysRevLett.126.097801} {\bibfield  {journal}
  {\bibinfo  {journal} {Phys. Rev. Lett.}\ }\textbf {\bibinfo {volume} {126}},\
  \bibinfo {pages} {097801} (\bibinfo {year} {2021})}\BibitemShut {NoStop}%
\bibitem [{\citenamefont {Prathyusha}\ \emph {et~al.}(2018)\citenamefont
  {Prathyusha}, \citenamefont {Henkes},\ and\ \citenamefont
  {Sknepnek}}]{Prathyusha}%
  \BibitemOpen
  \bibfield  {author} {\bibinfo {author} {\bibfnamefont {K.~R.}\ \bibnamefont
  {Prathyusha}}, \bibinfo {author} {\bibfnamefont {S.}~\bibnamefont {Henkes}},
  \ and\ \bibinfo {author} {\bibfnamefont {R.}~\bibnamefont {Sknepnek}},\
  }\href {\doibase 10.1103/PhysRevE.97.022606} {\bibfield  {journal} {\bibinfo
  {journal} {Phys. Rev. E}\ }\textbf {\bibinfo {volume} {97}},\ \bibinfo
  {pages} {022606} (\bibinfo {year} {2018})}\BibitemShut {NoStop}%
\end{thebibliography}%
   \end{document}